\documentclass[a4paper, 11pt]{article}
\usepackage{amssymb,amsmath}
\usepackage[T1]{fontenc} 
\usepackage[utf8]{inputenc}
\usepackage[english]{babel}
\usepackage{array}
\usepackage{booktabs} 	
\usepackage{caption}
\usepackage{graphicx}
\usepackage{subfig}	
\usepackage{authblk}
\usepackage{arydshln} 	
\usepackage{mathrsfs}	
\usepackage{multirow}	
\usepackage{siunitx}
\usepackage{graphicx,rotating,booktabs}
\usepackage{comment}

\usepackage[top=1.25in, bottom=1.75in,
left=1.25in, right=1.25in]{geometry}

\newcommand{\df}[1]{\ensuremath{\operatorname{d}\!{#1}}}
\newcommand*\diff[1]{\mathop{}\!\mathrm{d^#1}}
\newcommand{\Pnue}{\nu_{\mbox{\tiny e}}}
\newcommand{\APnue}{\overline{\nu}_{\!\;\!\mbox{\tiny e}}}
\newcommand{\Pnux}{\nu_{\mbox{\tiny $x$}}}
\newcommand\Tstrut{\rule{0pt}{2.6ex}}  
\newcommand\Bstrut{\rule[-1.1ex]{0pt}{0pt}}
\newcommand{\skpc}[1]{({#1})}

\DeclareSIUnit\erg{erg}
\DeclareSIUnit\foe{foe}
\DeclareSIUnit\parsec{pc}
\DeclareSIUnit\ton{ton}

\makeatletter
\newlength{\negph@wd}
\DeclareRobustCommand{\negphantom}[1]{%
  \ifmmode
    \mathpalette\negph@math{#1}%
  \else
    \negph@do{#1}%
  \fi
}
\newcommand{\negph@math}[2]{\negph@do{$\m@th#1#2$}}
\newcommand{\negph@do}[1]{%
  \settowidth{\negph@wd}{#1}%
  \hspace*{-\negph@wd}%
}
\makeatother

\newcommand{\vartilde}[1]{\mathpalette\dovartilde{#1}}
\newcommand{\dovartilde}[2]{%
\ifx#1\displaystyle\widetilde{#2}\else\tilde{#2}\fi
}

\usepackage{hyperref}
\hypersetup{
    bookmarks=true,         
    unicode=false,          
    pdftoolbar=true,        
    pdfmenubar=true,        
    pdffitwindow=false,     
    pdfstartview={FitH},    
    pdfauthor={Kevin Multani},     
    colorlinks=true,       
    linkcolor=blue,          
    citecolor=blue,        
}

\title{Late time supernova neutrino signal\\
and proto-neutron star radius}
\author[1,2,3]{A.\ Gallo Rosso\thanks{Corresponding
author: \href{mailto:andrea.gallorosso@gssi.it}
{andrea.gallorosso@gssi.it}.}}
\author[3]{S.\ Abbar\thanks{Corresponding
author: \href{abbar@apc.in2p3.fr}
{abbar@apc.in2p3.fr}.}}
\author[2]{F.\ Vissani}
\author[3]{M.C.\ Volpe}
\affil[1]{Gran Sasso Science Institute,
Viale F.\ Crispi 7, L'Aquila, Italy}
\affil[2]{INFN,
Laboratori Nazionali del Gran Sasso,
Via G. Acitelli, 22, Assergi, L'Aquila, Italy}
\affil[3]{Astro-Particule
et Cosmologie (APC),
CNRS UMR 7164, Universit\'e Denis Diderot\\
\hspace{\textwidth}
10, rue Alice Domon et L\'eonie Duquet,
75205 Paris Cedex 13, France}
\date{}
\begin{document}
\maketitle
\begin{abstract}
We discuss the possibility of
reconstructing the newly formed proto-neutron star
radius from the late time neutrino signal. 
A black-body emission is assumed for the neutron
star cooling phase. We parametrize the neutrino
time-integrated fluxes based on simulations
of Roberts and Reddy. A likelihood analysis of
the inverse-beta decay and elastic scattering
events in Hyper-Kamiokande is performed in both
three flavor and an effective one flavor scenario.
We show that the precision achievable in the radius
reconstruction strongly depends on a correlation
with the pinching parameter and therefore the
corresponding prior. 
Although this correlation hinders the precise
measurement of the newly formed neutron star
radius, it could help measure the pinching
parameters with good accuracy in view of the
current constraints on neutron star radius, or if
the neutron star radius is precisely measured.
\end{abstract}

\section{Introduction}
Neutrinos from core-collapse supernovae are an
incomparable laboratory for astroparticle physics.
The measurement of the neutrino luminosity curves
from future galactic explosions is fundamental
to fully unravel the explosion mechanism, to pin
down the newly born neutron-star or exotic neutrino
properties and to tell us about neutrino flavor
conversion in dense and explosive environments. 
The delayed explosion mechanism aided by the
standing accretion shock instability is thought to
be responsible for the vast majority of
core-collapse supernovae \cite{Bethe:1984ux}. 
Imprints of such an instability are expected to be
present in the neutrino time signal and be
detectable in IceCube \cite{Tamborra:2013laa}. In
addition, Colgate and Johnson hypothesized that the
gravitational binding energy of the newly formed
neutron star is taken away by electron, muon and
tau neutrinos and antineutrinos
\cite{Colgate:1966ax}. This conjecture, supported
by the analyses of SN1987A observations under the
equipartition assumption (see e.g.\
\cite{Loredo:2001rx,Pagliaroli:2009,%
Vissani:2014doa}) will be
precisely verified by the future measurement of the
gravitational binding energy of the newly formed
neutron star in Super-Kamiokande at $11 \%$
\cite{GalloRosso:2017hbp} and in Hyper-Kamiokande
at $3 \%$ percent level \cite{GalloRosso:2017mdz}. 

Concerning the exotic neutrino properties, neutrino
flavor transformations have been reliably
studied only on relatively small distances.
This circumstance does not allow us to exclude
firmly the possibility that new phenomena occur
when neutrinos travel on cosmic scales ---
thus leading to major modifications of the
supernova neutrino signal.
This was first discussed in the context
of the models in which the neutrino mass is
supposed to have pseudo-Dirac character, namely,
those models where left and right neutrinos exist
(see \cite{pd1,pd12}
and references therein).
Certain models for sterile neutrinos, based on the
hypothesis of exact mirror symmetry, have slightly
stronger theoretical and phenomenological bases.
They provide us, for instance, with convincing
candidates for dark matter particles (see e.g.\ 
\cite{m1}) and lead to very different masses for
neutrinos. However, they lead to the same
phenomenology for what concerns neutrino
oscillations. In fact, it was argued that a rather
interesting case for vacuum oscillations on cosmic
scales, is the one when the disappearance of the
ordinary neutrinos results in just  {\em half of
the original flux} in the case of mirror neutrinos
\cite{m2} or for pseudo-Dirac neutrinos \cite{pd2}.
The electron antineutrinos observed from SN~1987A
do not exclude this possibility strongly
\cite{m3}.

The neutrino signal from a future core-collapse
supernova will also provide us with information on
other macroscopic properties of the newly formed
neutron star, in particular its mass and radius,
just as in the case of SN~1987A 
\cite{Loredo:2001rx}.
Observationally, constraints on mass and radius 
of the neutron star, and consequently on its
equation of state can be obtained by Bayesian
analyses of quiescent low mass X-ray binaries
\cite{Lattimer:2013hma}. The NICER experiment will
measure the mass-radius relation as well as the
radius itself through the timing and the
spectroscopy of thermal and non-thermal emission 
 of neutron stars in the soft X-ray band
\cite{Gendreau}.
The expected sensitivity is at the level of $5 \%$
precision on the radii and will furnish tight
observational constraints on the neutron star 
equation of state. As for the neutron star masses,
they are precisely measured in radio binary pulsars
or in X-ray accreting binaries (see
\cite{Lattimer:2006xb} for a compilation of
measurements). In addition, gravitational waves
constitute a powerful probe for neutron star
properties and for extended theories of gravity.
The recent measurement of gravitational waves from
binary neutron stars has indeed yielded information
on the neutron star mass-radius relation and the
equation of state \cite{Abbott:2018exr}. 

A question one might ask is ``what would be the
prospects on measuring the radius of the nascent
neutron star in the future measurement of a
supernova neutrino time signal''. This
determination would rely on reference models for
the time signal, or the fluences. 
Moreover theoretical information would
be needed to establish the connection between the
neutrinosphere and the neutron star radii. 

In the present manuscript we address the issue
of the reconstruction of the neutron star radius
under the assumption of a black body emission
from the nascent neutron star. We consider 
neutrino fluences from the simulations
of Reddy and Roberts as a template.
In our
investigation, we study the signal observed in
Hyper-Kamiokande, the largest water Cherenkov
detector currently under consideration.
We combine inverse
beta decay and elastic scattering detection
channels and perform several nine-degrees of
freedom likelihood analyses. The total energy, the
average energy and the pinching parameters that
characterize the neutrino fluences for the three
neutrino species can vary within their
priors and are reconstructed using the simulated data.
We show that the inclusion of the pinching parameters
which quantify the deviations from the 
thermal distribution significantly affects the
results of the analysis.

We perform three analyses in which we
employ different priors for the pinching
parameter. Moreover, we consider a supernova
explosion at \SI{2}{\kilo\parsec} and at
\SI{10}{\kilo\parsec} as reference distances.
We discuss the precision in the
neutron star radius reconstruction and the
difficulties inherent to it. Finally, we
analyze the possibility to determine the pinching
parameter of the neutrino fluences by implementing
reasonable {\it ansatz} on the neutron star radius.  
 
The manuscript is structured as follows. In section
\ref{sec:due} we introduce the hypothesis assumed
for the neutrino emission from the nascent neutron
star, the time integrated neutrino fluxes and their
flavor modification. Section \ref{sec:ExpEvent}
presents the likelihood analyses performed in
Hyper-Kamiokande. Our numerical results concerning
the reconstruction of the neutron star radius from
the neutrino time signal is presented in section 
\ref{sec:tre}. The possibility to determine the
pinching parameter from the neutrino fluences is
also discussed. Section \ref{sec:quattro} is
devoted to conclusions.

\section{Neutrino time signal and proto-neutron
star cooling}
\label{sec:due}

\subsection{Parameterization of the
cooling phase time signal}
\label{sec:parametrizza}

During the proto-neutron star cooling phase,  
the neutrino emission can be approximately
described by a black-body of luminosity $L$
characterized by temperature $T$ and radius $R$
\begin{equation}
	L = 4 \pi \sigma_{\text{\textsc{bb}}} \phi(\eta)
	R^2\,k_B^4T^4,
	\label{eq:BBlaw0}
\end{equation}
where $\eta$ is the pinching parameter and $\sigma_{\text{\textsc{bb}}}$ is the black
body constant for a species with one degree of
freedom following a perfect Maxwell-Boltzmann
distribution
\begin{equation}
	\sigma_{\text{\textsc{bb}}} =
	\frac{45}{\pi^4}\times
	\frac{2\pi^5}{15 c^2 h^3} \approx
	\SI{4.75e35}
	{\erg\per\mega\electronvolt\tothe{4}
	\per\square\centi\meter\per\second}.
\end{equation}
The parameter $\phi$, of the order of unity,
accounts for deviations from such distribution.

Equation \eqref{eq:BBlaw0} can be used to
reconstruct the neutron star radius $R$, once the
neutrino flux parameters $L$, $T$, $\eta$ are
known. In the present work we consider as reference
the output of the detailed simulations by Roberts
and Reddy \cite{Roberts:2016rsf}. Such simulations
extend up to several seconds after core-bounce
where the neutrino emission is believed
to be related to the quasi-static Helmholtz cooling
of the proto-neutron star \cite{Janka:2006fh}.
Figure 3 of ref.\ \cite{Roberts:2016rsf} shows that
the time evolution of the flux parameters
and the neutrinosphere radii $R_{\nu,i}$,
are quite constant at later times for the three
neutrino species,
$\nu_{\mathrm{e}}$, $\overline{\nu}_{\mathrm{e}}$
and $\nu_x$.\footnote{Here $\nu_x$ indicates
$\nu_{\mu}$ or $\overline{\nu}_{\mu}$,
$\nu_{\tau}$, $\overline{\nu}_{\tau}$.}
Hence, in order to define a neutron-star radius we
consider the time window from \SI{6}{\second} to
\SI{10}{\second}. We note that the neutrinospheres
that depend both on the neutrino flavor and 
energy are defined as the location where the
opacity is equal to $2/3$ --- as in ref.\ 
\cite{Roberts:2016rsf}.
Since the neutron star radius is supposed to be
close to the neutrinosphere locations within
$10 \%$ \cite{Loredo:2001rx, Roberts:2016rsf},
in our analysis we consider the neutron star radius
to be at the same location as the neutrinosphere
radius.

The time-dependent (isotropic) neutrino fluxes at
the neutrinospheres $\Phi_i^0$ are given by
\begin{equation}
	\frac{\partial^2 \Phi_i^0}{\partial E
	\partial t} = \frac{\dot{N}_{\nu,i}^t}{4\pi{D}
	^{2}} f_i(E,t)
	\quad\text{with}\quad i = \nu_{\mathrm{e}},\,
	\overline{\nu}_{\mathrm{e}}\,\nu_x,
	\label{eq:flussoFD}
\end{equation}
where $\dot{N}_{\nu,i}^t = L^t_i/\langle E_i
\rangle^t$
is the number of emitted neutrinos per unit time,
the superscript $t$ recalls that we are dealing
with time-dependent quantities and $D$ is the
supernova distance. The $f_i(E,t)$ functions
parameterize  the energy distribution for each
species, normalized to 1. In the simulation
\cite{Roberts:2016rsf}, Fermi-Dirac distributions
are considered
\begin{equation}
	f_i\left(E,t\right) =
	\frac{E^2}{(k_B T_i^t)^3\,F_2(\eta_i^t)}
	\left[1+\exp\left(\frac{E}{k_B T_i^t}
		- \eta_i^t\right)\right]^{-1}
	\label{eq:distribFD}
\end{equation}
where the temperatures are related to the mean
energies $\langle E_i\rangle^t$ through
\begin{equation}
	k_B T_i^t = \frac{\left\langle
	E_i\right\rangle^t
	F_2(\eta_i^t)}{F_3(\eta_i^t)}.
	\label{eq:TedE}
\end{equation}
The $F_n\left(\eta\right)$ are the Fermi functions
\begin{equation}
	F_n(\eta) = \int_0^\infty \frac{x^n \df{x}}{1+
	\exp\left(x- \eta\right)} = - \,n!\:
	\mathrm{Li}_{n+1}\left(-e^\eta\right).
\end{equation}

From considerations of
statistical mechanics one can derive the expression
of the flux from a spherically symmetric source
emitting fermions --- see e.g.\ \cite{pathria1996}
\begin{equation}
	\label{eq:fluxStat}
	\frac{\df\Phi_\nu^{\text{\textsc{sm}}}}{\df E} 
	=\frac{\pi R^2 c}{4\pi D^2}
	\frac{4 \pi E^2}{c^3 h^3}
	\left[1 + \exp\left(\frac{E}{k_B T}-\eta
	\right)\right]^{-1}.
\end{equation}
Notice that, here $\eta$ is not related
to a chemical potential \cite{Dighe:1999bi}.
Comparing equations \eqref{eq:flussoFD} and
\eqref{eq:fluxStat} we obtain an explicit relation
that links the luminosity $L_i^t$, the radius
$R_i^t$, the temperature $T_i^t$ and the pinching 
$\eta_i^t$
\begin{equation}
	L_i^t =-\frac{24 \pi ^2 c}{\left(h c\right)^3}
	\left(R_i^t\right)^2\left(k_B T_i^t\right)^4
	\mathrm{Li}_4
	(-e^{\eta_i^t}).
	\label{eq:BBlaw}
\end{equation}
This expression constitutes the black-body law we
will consider for the neutrino emission. Note that
the effect of departures from an exactly thermal
emission are included in this formula by the 
corrections introduced by the pinching parameter.

In principle, the reconstruction of the
neutron-star radius would require 12 degree
time-dependent likelihood analysis --- $L_i^t$,
$\langle E\rangle^t$, $\eta_i^t$ and $R_i^t$,
satisfying equation\ \eqref{eq:BBlaw}. According to
the results of ref.\ \cite{Roberts:2016rsf}
(figure 3), the quantities $L_i^t$, $\langle E
\rangle^t$ and $R_i^t$ are quite constant in the
chosen time window. The pinching parameter can be
deduced from equation \eqref{eq:BBlaw} in terms of
the luminosity, average energy and radius at each
time.
Obviously, $\eta_i^t$ turns out to be approximately 
time independent as well. As a consequence, we do
not work with instantaneous fluxes but instead we
integrate equation \eqref{eq:flussoFD} over time to
obtain fluences. We will then be dealing with
effective $L_i$, $\langle E\rangle$ and $\eta$.
From these quantities and by using equation
\eqref{eq:BBlaw}, we reconstruct $R_i$ that will
also be an effective parameter.

Another justification to the time-independent
likelihood analysis we perform is that a precise
determination requires a large statistics, for
which the supernova location $D$ is a key
parameter. For the typical value
$D=\SI{10}{\kilo\parsec}$ corresponding to
the mean of the core-collapse supernova
distribution in our Galaxy \cite{Costantini:2005un}
and in a restricted time window for the neutrino
signal, the luminosities have already fainted
considerably. As we will show (section
\ref{sec:ExpEvent}), the number of expected events
in Hyper-Kamiokande reduces to few thousands in the 
$6\div\SI{10}{\second}$ time window,\footnote{The
same order of magnitude as the number expected for
the whole emission in Super-Kamiokande.} not
enough to be shared among time-bins preserving
a reasonable reconstruction accuracy (see
\cite{GalloRosso:2017mdz} for details).
As a term of comparison we will also present
results for a supernova at \SI{2}{\kilo\parsec} for
which the number of events raises though at several
tens of thousands. In conclusion, we will perform a
9 degree of freedom time-independent likelihood
analysis to determine the neutrino fluence
parameters and use equation \eqref{eq:BBlaw} to
reconstruct the effective neutron star radius.

\begin{table}[tbp]
	\centering
	\begin{tabular}{|l|rrr|l|}
		\hline
			& \multicolumn{1}{c}
			{$\nu_{\mathrm{e}}$\Tstrut}
			& \multicolumn{1}{c}{
			$\overline{\nu}_{\mathrm{e}}$\Bstrut}
			& \multicolumn{1}{c|}{$\nu_x$}
			&\multicolumn{1}{c|}{Prior}\\
			\hline
			$\mathcal{E}_i^*\:[10^{51}
			\:\mathrm{erg}]$\Tstrut
			& 2.486 	& 1.965		& 3.625
			& $\in[0.5,10]$\\
			$\langle E_i^* \rangle\:[\mathrm{MeV}]$
			& 8.806 	& 9.145 		& 9.509
			& $\in[4,14]$\\
			$\alpha_i^*$
			& 2.387		& 2.279		& 2.456
			& $\in[2.1,3.5]$ or $[2.1,2.6]$
			or $[2.27,2.47]$
			\Bstrut\\
			\hline
	\end{tabular}
	\caption{True values of the parameters
	used in the likelihood analysis, obtained
	by a fit on the time integrated fluxes,
	provided by \cite{Roberts:2016rsf}, in the
	time window $6\div\SI{10}{\second}$ and for
	the three neutrino species. The
	last column presents the priors in which
	the parameters are free to vary. The three
	priors on $\alpha$ correspond to the different
	analyses performed (see text).
	Note that the priors on $\mathcal{E}_i$ and 
	$\langle E_i \rangle$ are large enough to cover
	the whole region of the extracted points for
	the $\overline{\nu}_{\mathrm{e}}$ and $\nu_x$.
	}
 	\label{tab:param}
\end{table}

By integrating equation \eqref{eq:flussoFD}, the
fluence is
\begin{equation}
	\label{eq:fluenza}
	\frac{\mathrm{d} F_i^0}{\mathrm{d} E} = 
	\frac{N_{i,\nu}}{4\pi D^2} f_i(E) 
	\quad\text{with}\quad
	N_{i,\nu} =
	\frac{\mathcal{E}_i}{\langle E\rangle}.	
\end{equation}
where $\mathcal{E}_i$ is the total emitted energy
in the $i$-th species. Instead of using the
Fermi-Dirac distribution as in ref.\ 
\cite{Roberts:2016rsf}, we take for the energy
distribution $f_i(E)$  the ``Garching
parameterization'' \cite{Tamborra:2012ac}, as 
in our previous works \cite{GalloRosso:2017hbp,
GalloRosso:2017mdz}.
For each species the function $f_i(E)$ is given by
\begin{equation}
	f_i(E) = \frac{1}{\Gamma(\alpha_i +1)}
	\frac{E^{\alpha_i}}{T_i^{\alpha_i + 1}}
	e^{E/T_i}
	\quad\text{with}\quad
	T_i = \frac{\langle E_i \rangle}
	{(\alpha_i +1)}.
	\label{eq:distribAlfa}
\end{equation}
where, here, the pinching is expressed in terms of
the parameter $\alpha$. The two distributions
\eqref{eq:distribFD} and \eqref{eq:distribAlfa} are
strictly related, in the sense that it is possible
to link the respective pinching parameters $\eta$
and $\alpha$. Indeed one can define
\begin{equation}
	I_{n,k} = \int_0^{\infty} \df{E}\:
	E^n f_{k}(E) 
\end{equation}
with $k = FD, \alpha$ (for Fermi-Dirac and
Garching parameterization respectively)
and compare the width of the two distributions
with respect to their mean value
\begin{equation}
	\frac{\sqrt{I_{2,FD}-(I_{1,FD})^2}}{I_{1,FD}
	} = \frac{\sqrt{I_{2,\alpha}-
	(I_{1,\alpha})^2}}{I_{1,\alpha}}.
	\label{eq:campana}
\end{equation}
This gives a one-to-one map between $\eta$ and
$\alpha$, provided that $\alpha>2$. Note that the
value $\alpha = 2.30$ corresponds to an un-pinched
Fermi-Dirac distribution ($\eta = 0$). For
$\alpha\to 2$ one finds $\eta\to -\infty$.
Therefore the employment of the Garching
parameterization is more convenient than the
Fermi-Dirac one.
Combining equations \eqref{eq:fluenza} and
\eqref{eq:distribAlfa}, the complete expression
of the fluences become
\begin{equation}
	\frac{\mathrm{d} F_i^0}{\mathrm{d} E} = 
	\frac{\mathcal{E}_i}{4\pi D^2}
	\frac{E^{\alpha_i}\:
	e^{-E/T_i}}{T_i^{\alpha_i +2 }\:
	\Gamma\left(\alpha_i+2\right)}
	\quad\text{with}\quad i =
	\nu_{\mathrm{e}},\,
	\overline{\nu}_{\mathrm{e}}\,\nu_x.
	\label{eq:alfluenza}
\end{equation}

\begin{figure}[t]
	\centering
	\subfloat[][Fluences]
	{\includegraphics[width=.45\textwidth]
	{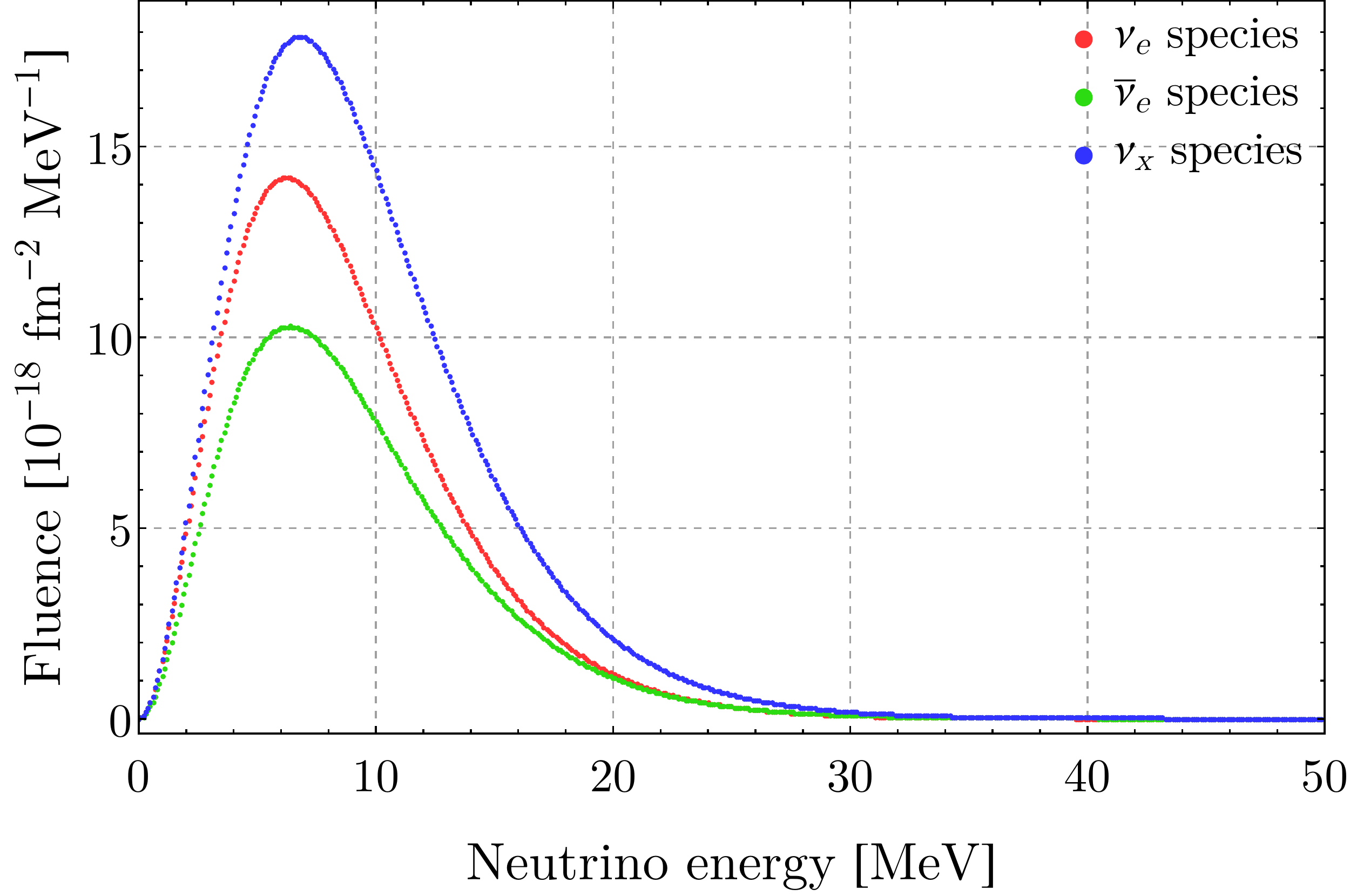}
	\label{fig:fluenza1}}
	\subfloat[][$\overline{\nu}_{\mathrm{e}}$
	fit function]
	{\includegraphics[width=.45\textwidth]
	{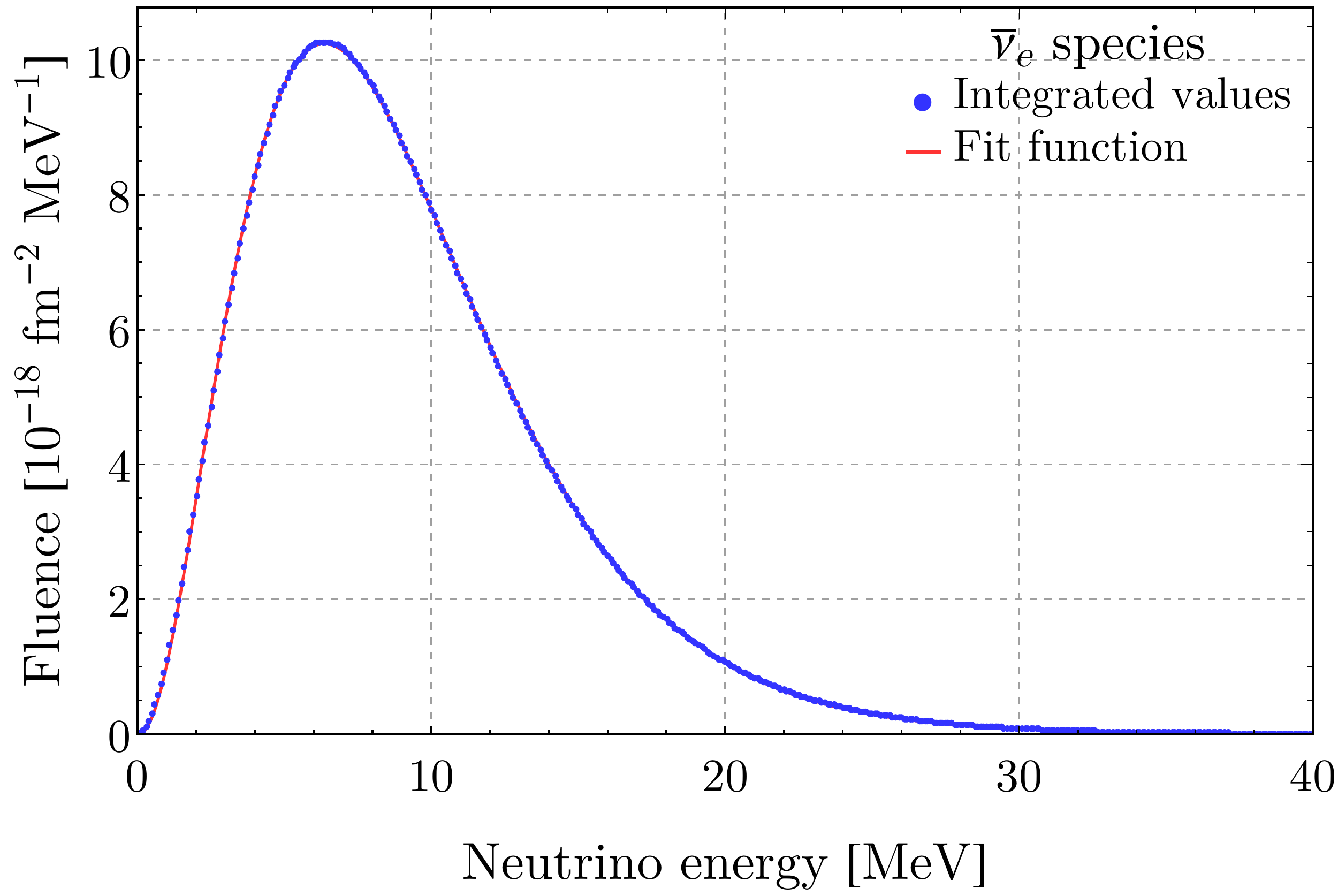}
	\label{fig:fluenza2}}
	\caption{\protect\subref{fig:fluenza1}
	Fluences for the three neutrino species
	obtained by time integration of equation
	\eqref{eq:flussoFD} with the parameters of
	figure 3 in ref.\ \cite{Roberts:2016rsf}.
	\protect\subref{fig:fluenza2} Comparison,
	for $\overline{\nu}_{\mathrm{e}}$,
	of the fluence shown in the left panel and
	the Garching energy distribution 
	\eqref{eq:alfluenza} with
	the effective parameters reported in table
	\ref{tab:param}.
	}
	\label{fig:influenze}
\end{figure}

The fluences \eqref{eq:alfluenza} are computed 
in the following way. First of all we use
expression \eqref{eq:flussoFD} with the results of
figure 3 of ref.\ \cite{Roberts:2016rsf} and
integrate it  in the window $6\div\SI{10}{\second}$
for each energy. The corresponding results are
provided in figure \ref{fig:fluenza1} for the three
neutrino species. Then, we perform a fit of such
results with the functional form given by equation
\eqref{eq:alfluenza} and extract the
time-independent effective parameters
$\mathcal{E}_i^*$, $\langle E_i^* \rangle$,
$\alpha_i^*$. These will be the true values of our
likelihood analysis (table \ref{tab:param}). 
Figure \ref{fig:fluenza2} shows as an example the
integrated distribution and the fitted one for 
$\overline{\nu}_{\mathrm{e}}$. As one can see,
the comparison of the integrated Fermi-Dirac fluxes
used in ref.\ \cite{Roberts:2016rsf} and
the Garching ones we employ is excellent.

\subsection{Flavor transformation of the
neutrino fluences}
During the neutrino propagation through the star,
the neutrino fluences can undergo spectral
swappings due to shock waves, turbulence and 
to the neutrino interactions with the
matter composing the astrophysical medium as well
as with background neutrinos and antineutrinos. 
Such interactions are implemented in the mean-field
neutrino evolution equations and produce a variety
of flavor conversion phenomena
\cite{Volpe:2016bkp}. Since the impact of 
neutrino self-interactions on the neutrino fluxes
still need to be fully assessed, here we only
consider the fluence modification due to the
Mikheyev-Smirnov-Wolfenstein (MSW) effect
\cite{Wolfenstein:1977ue,Mikheev:1986gs}, as done
in refs.\cite{GalloRosso:2017hbp,
GalloRosso:2017mdz}. We take normal ordering as
reference scenario. Therefore, the fluences at
detection $F_{\Pnue}, F_{\APnue} $ become the
following linear combination of the neutrino
fluences at the neutrinosphere
$F_{\APnue}^0, F_{\Pnux}^0$
\cite{Dighe:1999bi,Capozzi:2017ipn}:
\begin{equation}\label{eq:fluIBD}
	\begin{cases}
		F_{\Pnue} &=
		F_{\Pnux}^0\\
		F_{\APnue} &= 
		P_{\mathrm{e}}
		\cdot F_{\APnue}^0 + 
		(1-P_{\mathrm{e}}) \cdot F_{\Pnux}^0,
	\end{cases}
\end{equation}
where $P_{\mathrm{e}} = |U_{e1}|^2 =
|\cos\theta_{12} \cos\theta_{13}|^2 \approx 0.70$. 

\section{Likelihood analysis in Hyper-Kamiokande}
\label{sec:ExpEvent}

We study the signal given by a supernova explosion
located at a distance of $D=\SI{10}{\kilo\parsec}$,
assumed to be known precisely. To give a term of
comparison, in the following we will consider also
the optimistic situation of a supernova exploding
at $D=\SI{2}{\kilo\parsec}$.

We consider Hyper-Kamiokande as reference detector
which is expected to
have a fiducial mass of \SI{374}{\kilo\ton} and
to be built in the near future \cite{HK:2016dsw}.
We focus on the two main reactions of neutrino
detection  in water Cherenkov detectors, namely 
inverse beta decay (IBD) and elastic scattering
onto electrons (ES).

\begin{figure}[t]
	\centering
	\subfloat[][IBD events distribution]
	{\includegraphics[width=.45\textwidth]
	{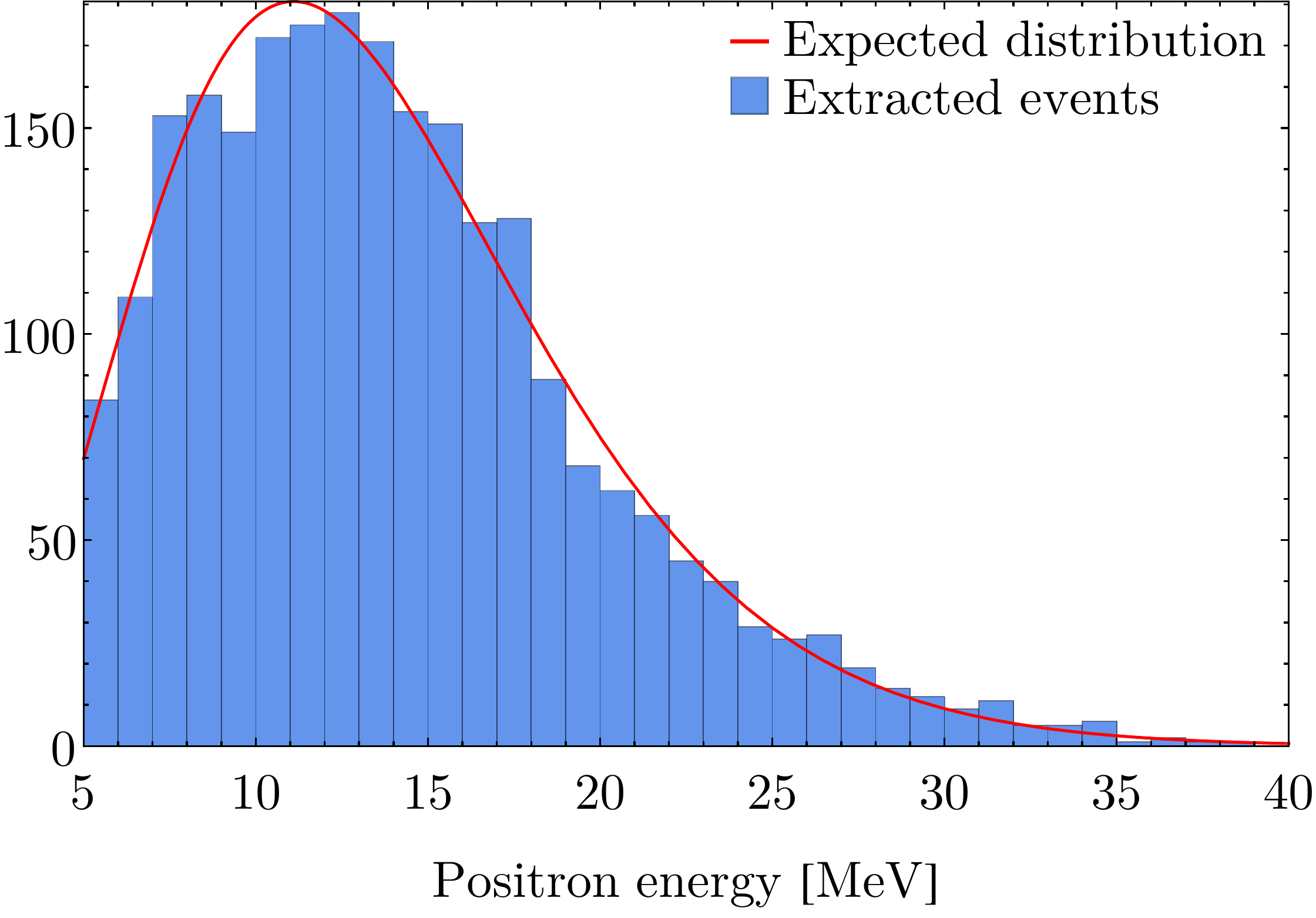}
	\label{fig:IBDeventi}}
	\subfloat[][ES events distribution]
	{\includegraphics[width=.45\textwidth]
	{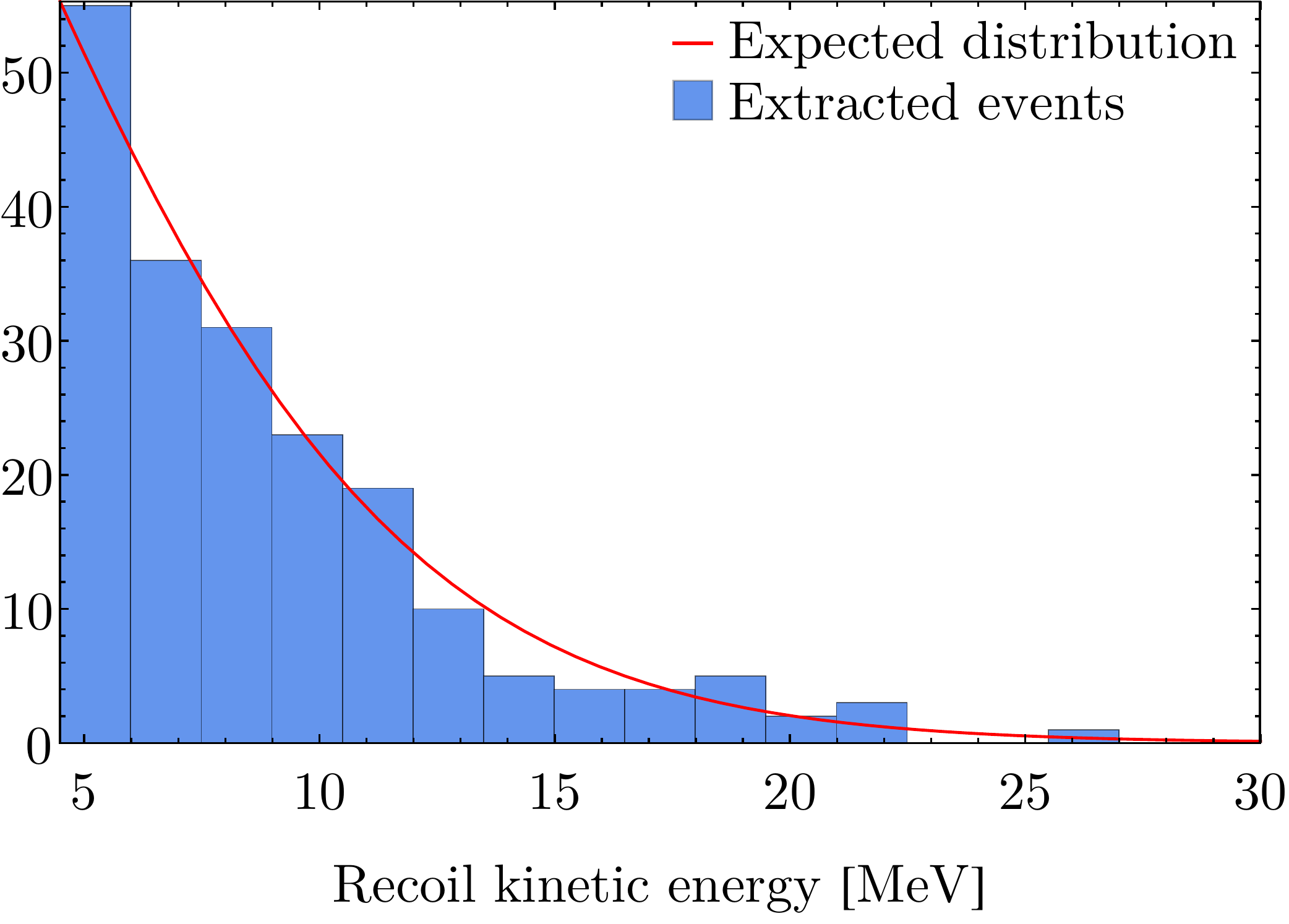}
	\label{fig:ESeventi}}
	\caption{
	Energy distribution of the 2437 IBD 
	\protect\subref{fig:IBDeventi}
	and 198 ES \protect\subref{fig:ESeventi}
	events. In each figure the histogram of the
	extracted data is plotted together with the
	expected distribution.
	}
	\label{fig:attese}
\end{figure}

Considering the process $ j = \mathrm{IBD}$, ES
and the neutrino species $i$, the number of
expected events can be expressed as
\begin{equation}
	\mathrm{N}_j = N_T \int_{E_{\mathrm{thr}}}
	^{\infty}
	\df{E}\,\frac{\df{F}_i}{\df{E}}\,\sigma_{ji},
\end{equation}
where $N_T$ is the number of targets and
$\sigma_{ji}$ is the cross section. The quantity
$E_{\mathrm{thr}}$ is the minimum of neutrino
energies which depends on the detector threshold
assumed to be \SI{5}{\mega\electronvolt} in the
following. The detailed expressions of
$\sigma_{ji}$ and $E_{\mathrm{thr}}$ as well as for 
the differential rates of expected events can be
found in ref.\  \cite{GalloRosso:2017mdz}. The
distributions of events are shown in figure
\ref{fig:attese}.

The total number of expected events with the
fluences, described by the true parameters in
table \ref{tab:param}, is
\begin{equation}
		\label{eq:Nattesi10}
		\mathrm{N}_{\mathrm{IBD}} = 2566\qquad
		\mathrm{N}_{\mathrm{ES}} = 188.
\end{equation}
A real experiment would see a number of events
$n_j$ given by Poissonian variation of the expected
ones. In our case we obtained
\begin{equation}\label{events}
		n_{\mathrm{IBD}} = 2437\qquad
		n_{\mathrm{ES}} = 198.
\end{equation}
For the \SI{2}{\kilo\parsec} case we
found $\mathrm{N}_{\mathrm{IBD}} = 64151$,
$\mathrm{N}_{\mathrm{ES}}= 4707$, $n_{\mathrm{IBD}}
= 64278$, $n_{\mathrm{ES}} = 4739$.

Following the theoretical distribution,
for each event we extract the corresponding
neutrino energy supposed to be known with
negligible smearing. The results for IBD and ES are
shown in figures \ref{fig:IBDeventi} and 
\ref{fig:ESeventi} respectively and compared with
the theoretical expectation.

In order to analyze the IBD and ES extracted
events we use a binned likelihood
\begin{equation}
\label{eq:LikBin}
	\mathcal{L}_{j}\left(\text{param.}
	\right)\propto
	\prod_{i=1}^{N_{\text{bin}}}
	\frac{\nu_i^{n_i}}{n_i} e^{-\nu_i}
	\quad\text{with}\:j=\text{IBD, ES};
\end{equation}
where the number of expected
events $\nu_i$, for the process $j$, in the
$i$-th bin, is compared with the number $n_i$
got from the random extraction.
In order to be able to get the most from the
extracted events, bin widths vary according to the
energy. We adopt the same energy bin sizes as in 
refs.\ \cite{Minakata:2008nc,
GalloRosso:2017hbp,GalloRosso:2017mdz}.

The analysis begins with the definition of a prior
in which the parameters are free to vary
(\tablename\ \ref{tab:param}). The ones for
the total energies $\mathcal{E}_i$ and mean
energies $\langle E_i\rangle$ are a little lower
than the ones of our previous analyses
\cite{GalloRosso:2017hbp,GalloRosso:2017mdz}. This
is justified, since we consider the explosion at
late times. Moreover, note that for the
$\overline{\nu}_{\mathrm{e}}$ and $\nu_x$ species
the priors are wide enough to fully include the set
of accepted points at $3\sigma$ confidence
level (see below). This is not true for the
$\nu_{\mathrm{e}}$ species; however, its parameters
are almost undetermined in this kind of detectors,
so a wider prior has no purpose (see e.g.\ the
discussion in \cite{GalloRosso:2017mdz}).

We emphasize that, given the low number of IBD$+$ES
events \eqref{events}, based on the analysis done
for Super-Kamiokande \cite{GalloRosso:2017mdz} it
is clear that the pinching parameters $\alpha_i$
will be undetermined for all species. Indeed, the
likelihood is not able to constraint their
variation in the neighborhood of the true values,
but allows them to span almost uniformly up to
values that are not of any usefulness. Thus,
defining the prior is equivalent to constrain these
parameters. Therefore it is of crucial importance
how the $\alpha$ prior is defined, since, as we
will see, the reconstruction of the emission radii
is strictly related to the pinching range.

In the following, we perform three different
analysis. The first one, called ``default'' takes
as prior on $\alpha$ the conservative range
$\left[2.1,\,3.5\right]$. It is opposed to a
scenario in which we make the hypothesis that
$\alpha$ is well constrained by supernova
simulations. It will be referred to
``$\alpha$--prior\texttt{+}'' and the
range considered is $\alpha\in\left[2.27,\,
3.47\right]$. In the third scenario, called simply
``$\alpha$--prior'', we assume a
mildly-constrained range between $2.1$ and $2.6$.
Notice that in our previous works
\cite{GalloRosso:2017hbp,GalloRosso:2017mdz} we
considered as reasonable interval the range
$[1.5,\,3.5]$, suggested in ref.\ 
\cite{Vissani:2014doa} for the time-independent
analysis including the whole neutrino time signal.
However, $\alpha\to 2$ approaches to a
Maxwell-Boltzmann distribution, that means $\eta\to
-\infty$ and consequently the loss of consistency
of equation \eqref{eq:BBlaw}.\footnote{It is also
unpractical to take $\alpha$ values very close to
2, since e.g.\ performing an analysis with $\alpha
=2 + 10^{-9}$ ($\eta=-19.7$) as lower limit would
lead to  huge tails in the neutron star radii that
would extend up to $\SI{2e4}{\kilo\meter}$ for the
$\nu_{\mathrm{e}}$, $\SI{8e3}{\kilo\meter}$ for the
$\overline{\nu}_{\mathrm{e}}$, and
$\SI{1.3e4}{\kilo\meter}$ for the $\nu_{x}$.}

Once the priors are defined (table
\ref{tab:param}), the Monte Carlo based analysis
is performed by extraction of random points inside
the $n$-dimensional region. Given a confidence
level (CL), each point $P$ is accepted if it
satisfies the relation
\begin{equation}\label{eq:chi}
	\log\mathcal{L}\left(P\right)\ge
	\log\mathcal{L}_{max} - \frac{A}{2},
	\quad\text{with}\quad
	\int_0^{A} \chi^2(N_{\mathrm{dof}}; z)
	\,\mathrm{d}{z} = \mathrm{CL};
\end{equation}
where $\chi^2(N_{\mathrm{dof}}; z)$ is a chi-square
distribution characterized by $N_{\mathrm{dof}}$
degrees of freedom and $\mathcal{L}_{max}$ is the
maximum of the (global) likelihood inside the
region. As done and discussed in our previous
papers, we assume a $100\%$ tagging efficiency
between IBD and ES events. In this manner, the
global likelihood is simply the product
between the IBD and ES ones.

\section{Reconstruction of the proto-neutron
star radius $R$}
\label{sec:tre}

We present here the numerical results of our
analysis. Two cases will be considered. In the
first one, the proto-neutron star radius is
reconstructed based on a three flavor framework and
a 9 degrees of freedom likelihood analysis. In the
second one we employ one effective neutrino flavor
and a three degrees of freedom likelihood analysis. 
We will not be discussing here the precision with
which one can determine the parameters defining
the neutrino fluences in eq.\ \eqref{eq:alfluenza}
--- see table \ref{tab:param}. Indeed this aspect
is investigated in detail in refs.\ 
\cite{GalloRosso:2017hbp,GalloRosso:2017mdz} for
the full neutrino time signal.\footnote{For the
interested reader, the precision in determining the
fluence parameters of present manuscript would
correspond to the full time signal in the
Super-Kamiokande case in
\cite{GalloRosso:2017mdz}.}

\subsection{Impact of the pinching parameter on
$R$ reconstruction}

\subsubsection{Three flavor analysis}

Once we find the accepted points according to
\eqref{eq:chi}, we can use the corresponding
parameters $\left(\mathcal{E}_i,\,
\langle E_i\rangle,\,\alpha_i\right)$ and
reconstruct the radii $R_i$ for the three species
by using equation \eqref{eq:BBlaw} and the relation
between $\alpha_i$ and $\eta_i$ \eqref{eq:campana}.
This gives us one point in the 12-dimensional
region following the parameters distribution.

Projecting these points onto the axes of the
multi-dimensional region gives PDF histograms,
the flux parameters $\mathcal{E}_i$, $\langle E_i
\rangle$, $\alpha_i$ for the three neutrino
species. Their mean and standard deviation
are reported in table \ref{tab:risu1} for the
default and $\alpha$--prior\text{+} analysis.

\begin{table}[t]
	\centering
	\begin{tabular}{|c|r|c|ccc|ccc|}
		\cmidrule{4-9}
		\multicolumn{3}{c|}{\Tstrut~\Bstrut}
		&\multicolumn{3}{c|}{\textsc{default}}
		&\multicolumn{3}{c|}{%
		$\alpha$--\textsc{prior}\texttt{+}}\\
		\cmidrule{3-9}
		\multicolumn{1}{c}{\Tstrut~}
		&\multicolumn{1}{c}{~\Bstrut}
		&\multicolumn{1}{|c|}{True}
		&\multicolumn{1}{c}{Mean}
		&\multicolumn{1}{c}{SD}
		&\multicolumn{1}{c|}{\%}
		&\multicolumn{1}{c}{Mean}
		&\multicolumn{1}{c}{SD}
		&\multicolumn{1}{c|}{\%}\\
		\hline
		\multirow{6}{*}{
		$\nu_{\mathrm{e}}$}
		&\multirow{2}{*}{$\mathcal{E}$ [foe]}
		&\multirow{2}{*}{2.49}
		&\Tstrut
		5.22 & 2.7 & 52.5 & 5.21 & 2.7 & 52.7\\
		& & & \skpc{5.10} & \skpc{2.7}
		& \skpc{53.5}
		& \skpc{4.68} & \skpc{2.7} & \skpc{57.4}
		\Bstrut\\
		\cdashline{3-9}
		&\multirow{2}{*}{$\langle E\rangle$ [MeV]}
		&\multirow{2}{*}{8.81}
		&\Tstrut
		8.95 & 2.9 & 32.0 & 8.93 & 2.9 & 32.0\\
		& & & \skpc{9.15} & \skpc{2.7}
		& \skpc{29.0} & \skpc{8.7} & \skpc{2.5}
		& \skpc{29.0}\Bstrut\\
		\cdashline{3-9}
		&\multirow{2}{*}{$\alpha$}
		&\multirow{2}{*}{2.39}
		&\Tstrut
		2.80 & 0.40 & 14.4 & 2.37 & 0.06 & 2.43\\
		& & & \skpc{2.82} & \skpc{0.40}
		& \skpc{14.3} & \skpc{2.37} & 
		\skpc{0.06} & \skpc{2.43}
		\Bstrut\\
		\hline
		\multirow{6}{*}{
		$\overline{\nu}_{\mathrm{e}}$}
		&\multirow{2}{*}{$\mathcal{E}$ [foe]}
		&\multirow{2}{*}{1.96}
		&\Tstrut
		1.66 & 0.39 & 23.4 & 1.71 & 0.41 & 23.7\\
		&&& \skpc{2.04} & \skpc{0.17} & \skpc{8.34}
		& \skpc{2.02} & \skpc{0.16} & \skpc{8.07}
		\Bstrut\\
		\cdashline{3-9}
		&\multirow{2}{*}{$\langle E\rangle$ [MeV]}
		&\multirow{2}{*}{9.14}
		&\Tstrut
		9.78 & 1.2 & 11.8 & 9.38 & 0.82 & 8.79\\
		&&& \skpc{9.35} & \skpc{0.44} & \skpc{4.69}
		& \skpc{9.32} & \skpc{0.29} &\skpc{3.13}
		\Bstrut\\
		\cdashline{3-9}
		&\multirow{2}{*}{$\alpha$}
		&\multirow{2}{*}{2.28}
		&\Tstrut
		2.78 & 0.4 & 14.3 & 2.37 & 0.06 & 2.44\\
		&&& \skpc{2.38} & \skpc{0.2} & \skpc{8.32}
		& \skpc{2.37} & \skpc{0.06} &\skpc{2.41}
		\Bstrut\\
		\hline
		\multirow{6}{*}{
		$\nu_{x}$}
		&\multirow{2}{*}{$\mathcal{E}$ [foe]}
		&\multirow{2}{*}{3.62}
		&\Tstrut
		3.78 & 0.81 & 21.3 & 3.85 & 0.84 & 21.8\\
		&&& \skpc{3.43} & \skpc{0.38} & \skpc{11.}
		& \skpc{3.52} & \skpc{0.36} & \skpc{10.1}
		\Bstrut\\
		\cdashline{3-9}
		&\multirow{2}{*}{$\langle E\rangle$ [MeV]}
		&\multirow{2}{*}{9.51}
		&\Tstrut
		9.71 & 1.1 & 11.4 & 9.31 & 0.8 & 8.62\\
		&&& \skpc{9.45} & \skpc{0.57} & \skpc{5.99}
		& \skpc{9.33} & \skpc{0.38} & \skpc{4.02}
		\Bstrut\\
		\cdashline{3-9}
		&\multirow{2}{*}{$\alpha$}
		&\multirow{2}{*}{2.46}
		&\Tstrut
		2.77 & 0.4 & 14.3 & 2.37 & 0.06 & 2.43\\
		&&& \skpc{2.52} & \skpc{0.29} & \skpc{11.4}
		& \skpc{2.37} & \skpc{0.06} &\skpc{2.42}
		\Bstrut\\
		\hline
	\end{tabular}
	\caption{Results of the flux parameter
	reconstruction in Hyper-Kamiokande, for the
	three neutrino species,	from the time-window
	$6\div\SI{10}{\second}$, combining
	IBD and ES signal. The two blocks named
	``default'' and ``$\alpha$--prior\texttt{+}''
	refer to an $\alpha$ prior of $[2.1,\,3.5]$
	and $[2.27,\,2.47]$ respectively. The top rows
	present the $D=\SI{10}{\kilo\parsec}$
	analysis, while the quantities in parentheses
	are the result of a $D=\SI{2}{\kilo\parsec}$
	explosion. For each parameter we give its
	true value,	the mean
	of the parameter distribution, its standard 
	deviation (SD) and the overall accuracy.}
	\label{tab:risu1}
\end{table}

\begin{figure}[t]
	\centering
	\subfloat[][Default case]
	{\includegraphics[width=.45\textwidth]
	{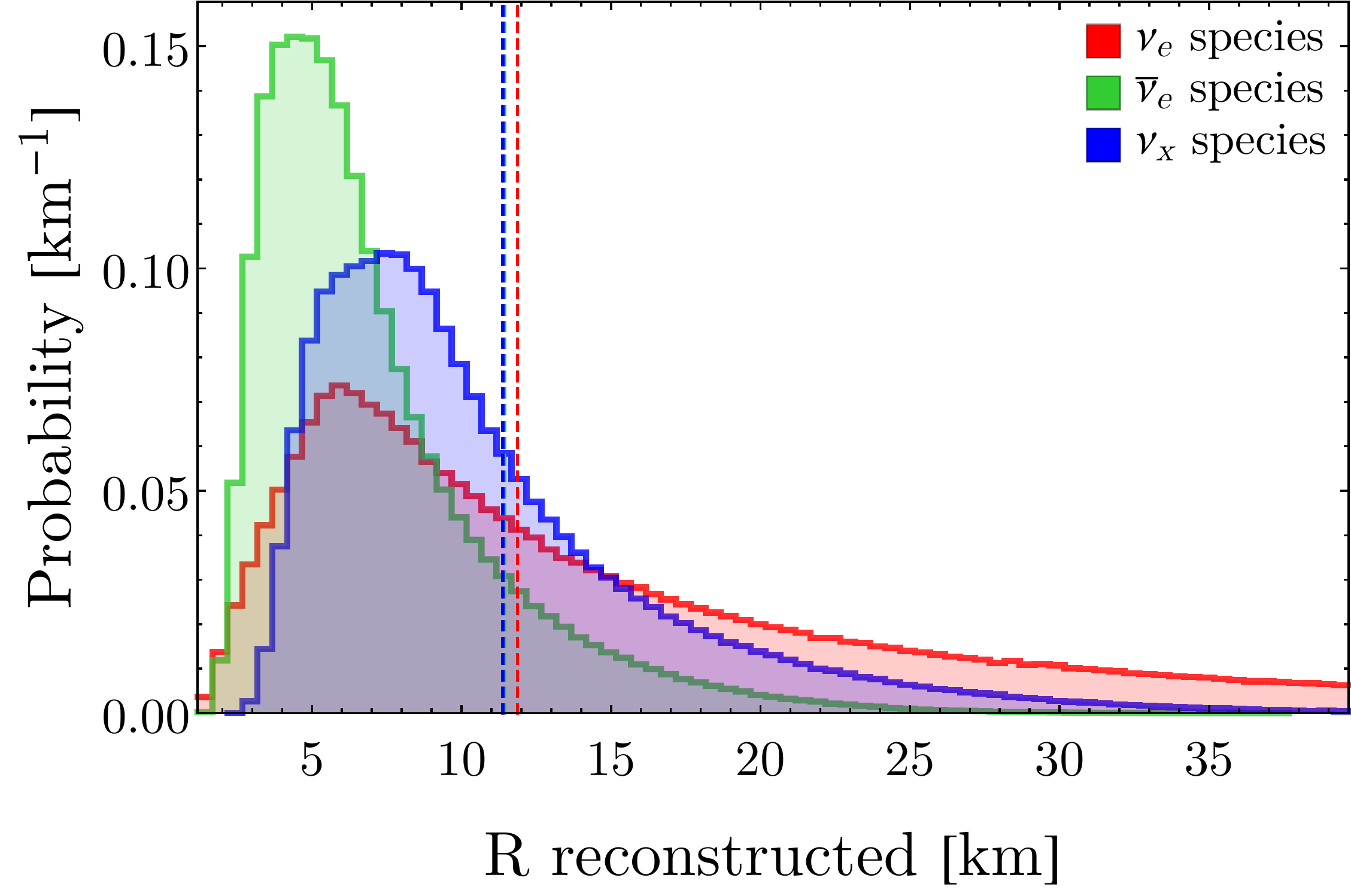}
	\label{fig:HR3a0}}
	\subfloat[][{$\alpha$--prior\texttt{+}
	case}]
	{\includegraphics[width=.45\textwidth]
	{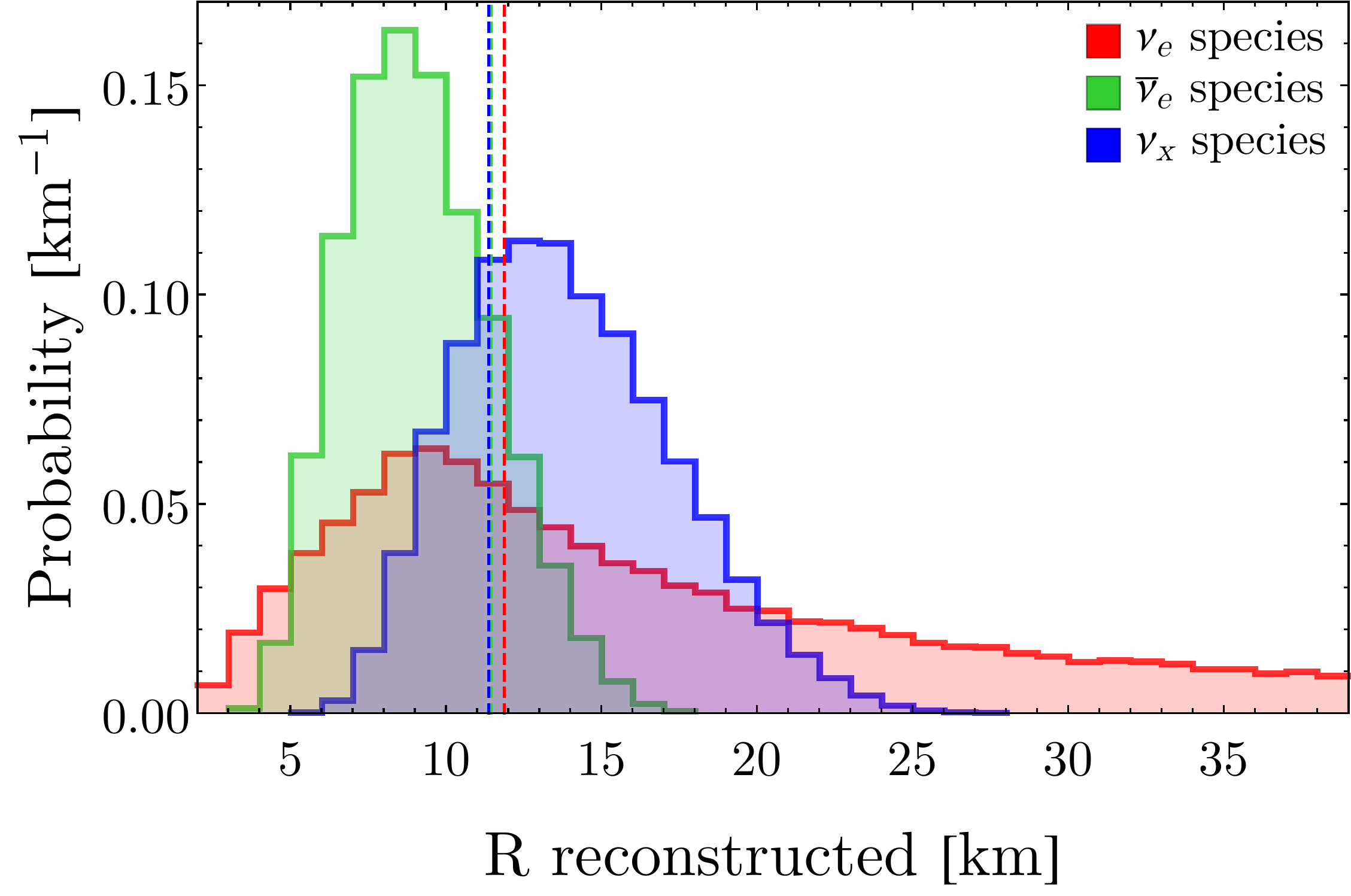}
	\label{fig:HR3a1}}
	\caption{
	Distribution of the reconstructed proto-neutron
	star radii for the three neutrino species. 
	The left panel \protect\subref{fig:HR3a0} shows
	the results obtained with the most conservative
	prior, while the right panel
	\protect\subref{fig:HR3a1} presents the ones
	for a well defined pinching parameter (see
	text). The dashed vertical lines show the true
	values for the radii. Notice that the radius
	tail for the $\nu_{\mathrm{e}}$ species extends
	up to \SI{214}{\kilo\meter} and \SI{130}
	{\kilo\meter} in figure
	\protect\subref{fig:HR3a0} and
	\protect\subref{fig:HR3a1} respectively.	}
	\label{fig:HistoR3}
\end{figure}

Looking at the $\SI{10}{\kilo\parsec}$ case in 
table \ref{tab:risu1}, the total energy
$\mathcal{E}_{\APnue}$ and the mean energy $\langle
E_{\APnue}\rangle$ are reconstructed with an
accuracy of $\sim 23\%$ and $\sim 12\%$.
The situation does not improve much even
considering the very aggressive scenario
``$\alpha$--prior\texttt{+}''. The knowledge of the
$\nu_x$ reaches an accuracy on
$\mathcal{E}_{\Pnux}$ and $\langle
E_{\Pnux}\rangle$ of $\sim 21\%$ and $\sim 11\%$
respectively. Except for the pinching, the accuracy
in the reconstruction of the total and average
energies for $\overline{\nu}_{\mathrm{e}}$ and
$\nu_x$ improves for the \SI{2}{\kilo\parsec}
case.

The pinching parameters $\alpha_i$ are
reconstructed with almost the same accuracies among
species, namely $\sim 14\%$ and $\sim 2.4\%$ for
the ``default''and ``$\alpha$--prior\texttt{+}''
analyses respectively. In the latter case the
accuracy obviously improves but only because the
prior is tighter. Indeed, the listed standard
deviations are similar to the ones expected for a
flat distribution inside the range of the prior, 
namely
$0.40$ for $\alpha\in\left[2.1,\,3.5\right]$,
$0.14$ for $\alpha\in\left[2.1,\,2.6\right]$ and
$0.06$ for $\alpha\in\left[2.27,\,2.47\right]$.
This is the same reason of the ``better''
reconstruction of the $\nu_{\mathrm{e}}$ species:
it is just a matter of prior, being the flux
properties of this species almost undetermined.

The neutrino signal in Super-Kamiokande and
Hyper-Kamiokande have been already analyzed in 
\cite{GalloRosso:2017mdz}. It is worthwhile to
compare these cases with the signal in the time
window $6\div\SI{10}{\second}$.
The number of detected events in Hyper-Kamiokande
in the window $6\div\SI{10}{\second}$ is similar to
the number of events seen for the whole
explosion in Super-Kamiokande at the same distance.
The total number of expected events in
Hyper-kamiokande, in the late-time window, for an
explosion at $D=\SI{2}{\kilo\parsec}$ is comparable
with the total number of events it would see at
$D=\SI{10}{\kilo\parsec}$. Therefore, one might
naively expect to reconstruct the quantities
$\mathcal{E}_i$ and $\langle E_i\rangle$ with the
same accuracies discussed in ref.\ 
\cite{GalloRosso:2017mdz} for the whole signal in
Super-Kamiokande and Hyper-Kamiokande
at $D=\SI{10}{\kilo\parsec}$. However, this is not
the case. In particular, concerning the
$\overline{\nu}_{\mathrm{e}}$ species,
the reconstruction of the total energy
$\mathcal{E}_{\APnue}$ and the mean energy $\langle
E_{\APnue}\rangle$ worsens by a factor of 2.
These features are not due to the likelihoods or
the particular analyses: redoing the
calculations using a likelihood with smaller bins
lead to the same results. A possible explanation of
this behavior may lie in the different values of
the distributions. Indeed, the total emitted
energies $\mathcal{E}_i$ and mean energies $\langle
E_i\rangle$ of the late time emission are
significantly lower than the ones in the whole time
window \cite{GalloRosso:2017mdz}.

In all the analyses, the most significant
contribution to the statistics is
given by the IBD events. As already underlined in
\cite{Minakata:2008nc}, if the spectral shape
(pinching) is unknown, the oscillation mechanism
introduces a degeneracy in the fluxes, especially
if the IBD signal only is considered.
It has been shown in \cite{GalloRosso:2017hbp} that
the combination of inverse beta decay and elastic
scattering can break the degeneracy between the
total and the mean energies. However, it is
reasonable to expect this works better when the
contamination of $\nu_x$ is smaller. Among the
$2566$ expected IBD events at \SI{10}{\kilo\parsec}
\eqref{eq:Nattesi10}, $1393$ are due to the emitted
$\overline{\nu}_{\mathrm{e}}$ and $1173$ to the
emitted $\nu_x$, i.e.\ $54\%$ and $46\%$
respectively. With the true parameters assumed in
our previous papers \cite{GalloRosso:2017hbp,
GalloRosso:2017mdz} these percentage were $63\%$
and $37\%$. This suggest the previous analyses gave
better results because the two flux components were
``less entangled''. However, such a combination had
not helped to get a good identification of the
second moment of the fluence distribution, as shown
in ref.\ \cite{GalloRosso:2017mdz}.

These uncertainties propagate to $R_i$
reconstruction. The histograms describing the
radii distributions
given in the default and $\alpha$--prior\texttt{+}
analysis are shown in figure \ref{fig:HR3a0} and
\ref{fig:HR3a1} respectively, while the numerical
values are listed in table \ref{tab:raggi}.
As expected, in both cases $R_{\Pnue}$ is
undetermined, since the flux parameters for this
neutrino species remain unknown.
Concerning $R_{\APnue}$ and $R_{\Pnux}$,
the accuracy remains poor, being around $50\%$ in
the conservative range
$\alpha\in\left[2.1,\,3.5\right]$ and $25\%$ with
the tightest prior $\alpha\in\left[2.27,\,2.47
\right]$ at $3\sigma$ (CL).

\begin{table}[t]
	\centering
	\begin{tabular}{|c|c|ccc|ccc|ccc|}
		\cmidrule{3-11}
		\multicolumn{1}{c}{~}
		&\multicolumn{1}{c|}{\Tstrut~\Bstrut}
		&\multicolumn{3}{c|}{\textsc{default}}
		&\multicolumn{3}{c|}{%
		$\alpha$--\textsc{prior}}
		&\multicolumn{3}{c|}{%
		$\alpha$--\textsc{prior}\texttt{+}}\\
		\cmidrule{2-11}
		\multicolumn{1}{c|}{\Tstrut~}
		&\multicolumn{1}{c|}{true\Bstrut}
		&\multicolumn{1}{c}{Mean}
		&\multicolumn{1}{c}{SD}
		&\multicolumn{1}{c|}{Acc}
		&\multicolumn{1}{c}{Mean}
		&\multicolumn{1}{c}{SD}
		&\multicolumn{1}{c|}{Acc}
		&\multicolumn{1}{c}{Mean}
		&\multicolumn{1}{c}{SD}
		&\multicolumn{1}{c|}{Acc}\\
		\multicolumn{1}{c|}{\Tstrut~}
		&\multicolumn{1}{c|}{$[\si{\kilo\meter}]$}
		&\multicolumn{1}{c}{$[\si{\kilo\meter}]$}
		&\multicolumn{1}{c}{$[\si{\kilo\meter}]$}
		&\multicolumn{1}{c|}{$[\%]$}
		&\multicolumn{1}{c}{$[\si{\kilo\meter}]$}
		&\multicolumn{1}{c}{$[\si{\kilo\meter}]$}
		&\multicolumn{1}{c|}{$[\%]$}
		&\multicolumn{1}{c}{$[\si{\kilo\meter}]$}
		&\multicolumn{1}{c}{$[\si{\kilo\meter}]$}
		&\multicolumn{1}{c|}{$[\%]$}\\
		\hline
		\multirow{2}{*}{$\nu_{\mathrm{e}}$}
		& \multirow{2}{*}{11.9}
		& 18.9 & 18.6 & 98.0 & 26.4 & 24.1 & 91.3
		& 23.8 & 20.5 & 86.1\\
		& & \skpc{16.8} & \skpc{16.2} & \skpc{96.5}
		& \skpc{24.4} & \skpc{21.1} & \skpc{86.4}
		& \skpc{22.2} & \skpc{17.9} & \skpc{80.8}\\
		\hline
		\multirow{2}{*}{$\overline{\nu}_
		{\mathrm{e}}$}
		& \multirow{2}{*}{11.5}
		& 7.1 & 4.0 & 56.0 & 10.1 & 3.9 & 38.4
		& 9.2 & 2.4 & 25.6\\
		&&\skpc{10.9} & \skpc{3.8} & \skpc{34.4}
		& \skpc{10.8} & \skpc{3.1} & \skpc{28.7}
		& \skpc{9.9} & \skpc{1.2} & \skpc{12.2}\\
		\hline
		\multirow{2}{*}{$\nu_x$}
		& \multirow{2}{*}{11.4}
		& 10.8 & 5.9 & 54.6 & 15.4 & 5.8 & 37.4
		& 13.9 & 3.4 & 24.5\\
		&& \skpc{12.3} & \skpc{4.9} & \skpc{40.0}
		& \skpc{14.1} & \skpc{4.2} & \skpc{29.8}
		& \skpc{13.0} & \skpc{1.8} & \skpc{13.7}\\
		\hline
	\end{tabular}
	\caption{Results of the radii
	reconstruction in Hyper-Kamiokande, for the
	three neutrino species,	from the
	time-window $6\div\SI{10}{\second}$, combining
	IBD and ES signal. The three blocks named
	``default'', ``$\alpha$--prior'' and
	``$\alpha$--prior\texttt{+}'' refer to an
	$\alpha$ prior of $[2.1,\,3.5]$, $[2.1,\,2.6]$,
	and $[2.27,\,2.47]$ respectively. The top rows
	present the $D=\SI{10}{\kilo\parsec}$
	analysis, while the quantities in parentheses
	are the result of a $D=\SI{2}{\kilo\parsec}$
	explosion. For each radius we give its
	true value, the mean
	of the parameter distribution, its standard 
	deviation (SD) and the overall accuracy.}
	\label{tab:raggi}
\end{table}

\begin{figure}[t]
	\centering
	\subfloat[][$\overline{\nu}_{
	\mathrm{e}}$ species]
	{\includegraphics[width=.45\textwidth]
	{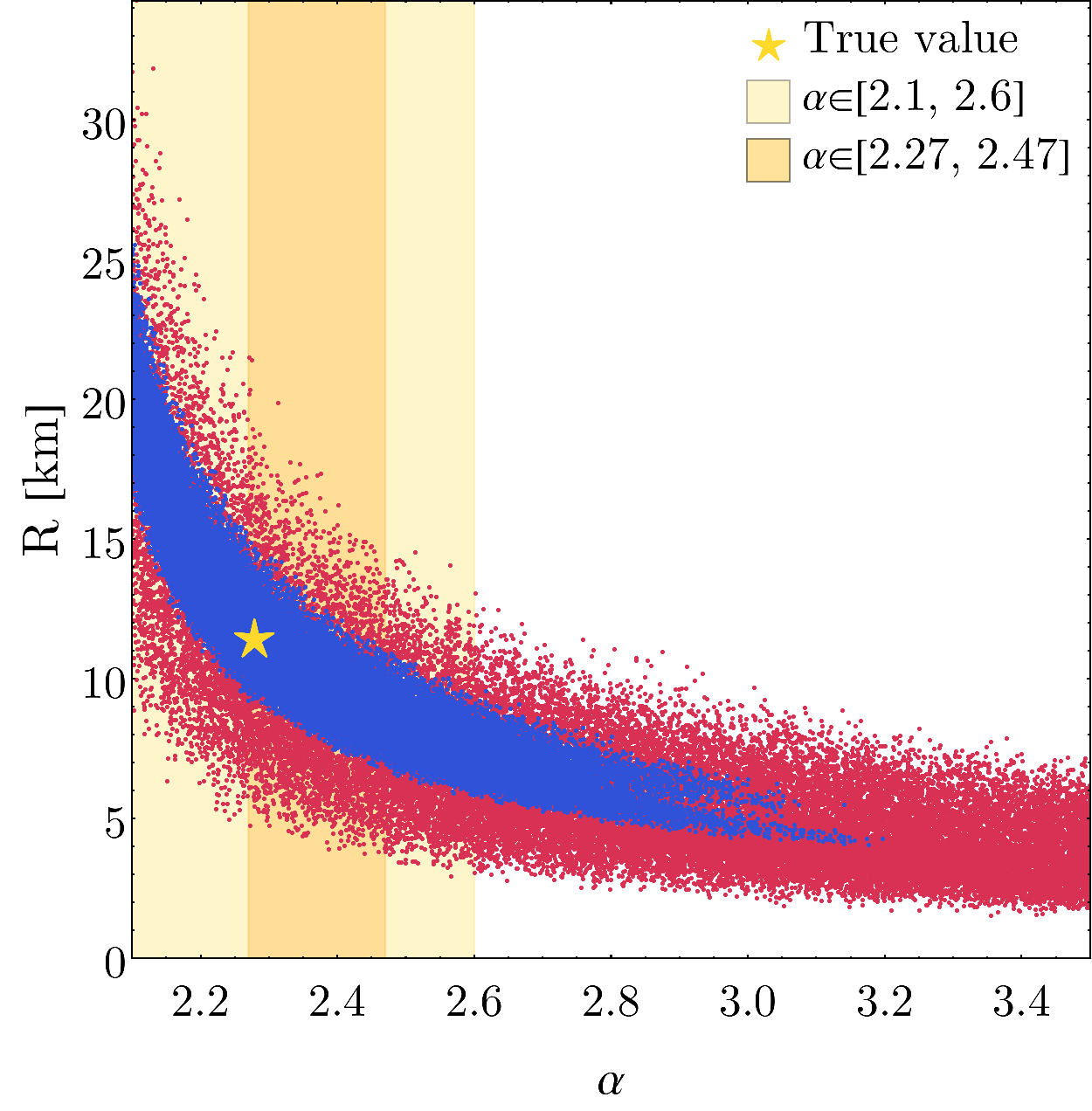}
	\label{fig:S3s2}}
	\subfloat[][$\nu_{x}$ species]
	{\includegraphics[width=.45\textwidth]
	{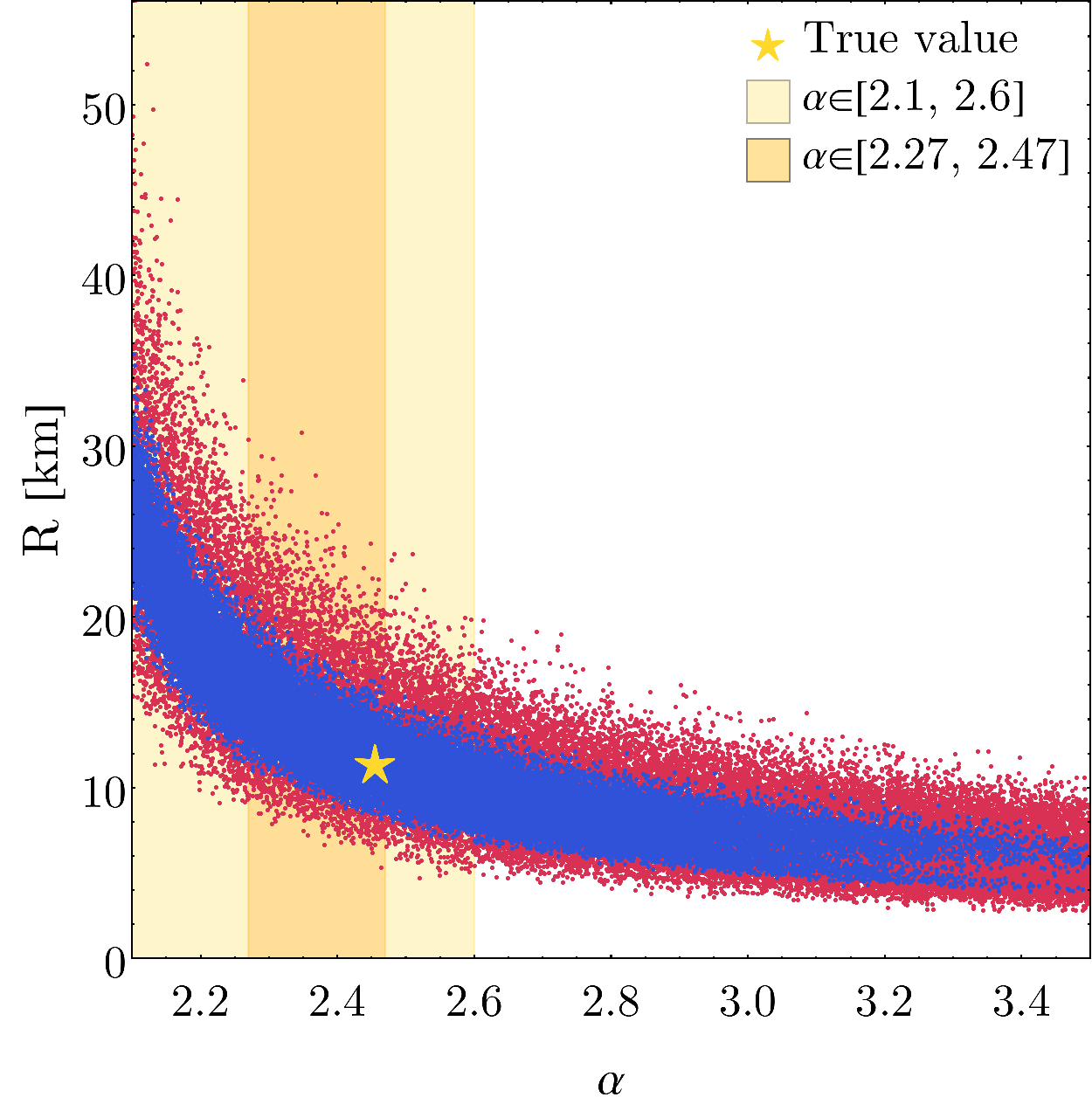}
	\label{fig:S3s3}}
	\caption{Projection onto the
	$\alpha$--$R$ plane of 40k points, in the
	$D=\SI{10}{\kilo\parsec}$ (red) and 
	$D=\SI{2}{\kilo\parsec}$ (blue) analysis,
	for the $\overline{\nu}_{\mathrm{e}}$
	\protect\subref{fig:S3s2} and $\nu_{x}$
	\protect\subref{fig:S3s3} species. In both
	panels, the star marks the true values, while
	the bands highlight the different assumptions
	on the prior of $\alpha$.}
	\label{fig:Scatter3}
\end{figure}

The explanation of this results can be found
projecting the extracted points onto
two-dimensional planes, as shown in figure
\ref{fig:Scatter3} for the $\alpha$--$R$ plane and
for the $\overline{\nu}_{\mathrm{e}}$ and
$\overline{\nu}_{x}$ species. From these plots, one
can clearly see a strong correlation between
$\alpha$ and $R$. This correlation does not
improve from the $D=\SI{10}{\kilo\parsec}$ to the 
$D=\SI{2}{\kilo\parsec}$ analysis, so it is not
due to the statistics. Indeed, it is somehow
expected. As we already discussed, the pinching
parameters are almost undetermined. The information
about their values, however, is crucial in the
radii reconstruction --- see equation 
\eqref{eq:BBlaw}. Indeed, figure \ref{fig:Scatter3}
shows a tight correlation that cannot be resolved
with neutrinos alone.

Moreover, notice that our ``true radii'' are
computed from equations (\ref{eq:BBlaw},
\ref{eq:campana}) by taking the true
values $\left(\mathcal{E}_i^*,\,
\langle E_i^*\rangle,\, \alpha_i^*\right)$ given in
table \ref{tab:param}. The result is reported in
the first column of table \ref{tab:raggi}.
We would like to underline that the values obtained
in this way are in principle different from the
radii computed in other ways, e.g.\ by the temporal
mean of the $R_i$ provided by the simulation of
ref.\ \cite{Roberts:2016rsf} and defined as the
region in which the neutrino opacity reaches
the value of $2/3$. The latter are
\begin{equation}
	\label{eq:raggimedi}
	R_{\Pnue}  = \SI{12.69}{\kilo\meter},\quad
	R_{\APnue} = \SI{12.19}{\kilo\meter},\quad
	R_{\Pnux}  = \SI{12.09}{\kilo\meter}.
\end{equation}
As one can see comparing table \ref{tab:risu1} and
equation \eqref{eq:raggimedi}, 
the latter are about $6\div 7\%$ bigger.

We remind that the radii reconstructed by the
neutrino data analyses are
neutrinosphere radii, that, according to
\cite{Roberts:2016rsf} are 
underneath the proto-neutron star radii in which we
are interested, while the two radii are the
same within $10 \%$ \cite{Loredo:2001rx,
Roberts:2016rsf}. In view of the importance of
this connection, a systematic work on simulations
is necessary, to quantify accurately the
comparison. When drawing conclusions on the
precision achieved for the proton-neutron star
radii, the entailed systematic 
errors should be taken into account.

\subsubsection{One effective flavor analysis}

\begin{table}[t]
	\centering
	\begin{tabular}{|r|c|ccc|ccc|}
		\cmidrule{3-8}
		\multicolumn{2}{c|}{\Tstrut~\Bstrut}
		&\multicolumn{3}{c|}{\textsc{default}}
		&\multicolumn{3}{c|}{%
		$\alpha$--\textsc{prior}\texttt{+}}\\
		\cmidrule{2-8}
		\multicolumn{1}{c|}{\Tstrut\Bstrut~}
		&\multicolumn{1}{c|}{True}
		&\multicolumn{1}{c}{Mean}
		&\multicolumn{1}{c}{SD}
		&\multicolumn{1}{c|}{Acc}
		&\multicolumn{1}{c}{Mean}
		&\multicolumn{1}{c}{SD}
		&\multicolumn{1}{c|}{Acc}\\
		\hline
		$\vartilde{\mathcal{E}}$ [foe]
		& 2.47\Tstrut
		& 2.31 & 0.11 & 4.6 & 2.33 & 0.1 & 4.41\\
		$\langle\vartilde{E}\rangle$ [MeV]& 9.31
		& 9.53 & 0.33 & 3.46 & 9.38 & 0.18 & 1.88\\
		$\vartilde{\alpha}$ & 2.35
		& 2.51 & 0.26 & 10.3 & 2.37 & 0.058 &2.43\\
		$\vartilde{R}$ [km] & 11.1
		& 9.9 & 3.8 & 38.3 & 10.5 & 1.1 & 10.6
		\Bstrut\\
		\hline
	\end{tabular}
	\caption{Results at $3\sigma$ CL as provided by
	the analysis on the IBD signal fitted with only
	one effective flavor $\vartilde{\nu}$. The two
	blocks named ``default'' and
	``$\alpha$--prior\texttt{+}''
	refer to an $\alpha$ prior of $[2.1,\,3.5]$,
	and $[2.27,\,2.47]$ respectively. For each
	parameter we listed the true value, mean,
	standard deviation (SD) and accuracy.}
	\label{tab:risu2}
\end{table}

One way to approach the issue of the
proto-neutron star radius reconstruction is to
perform an analysis based on one effective flavor
$\vartilde{\nu}$, instead of three
flavors. In this case, the expected IBD
flux\footnote{Namely, the one that gives the
highest statistics.} in figure \ref{fig:IBDeventi}
can be fitted assuming a Garching function
depending upon three parameters only, whose true
values are
\begin{equation}
	\vartilde{\mathcal{E}}^* =
	\SI{2.4727}{\foe};\quad
	\langle\vartilde{E^*}\rangle =
	\SI{9.3065}{\mega\electronvolt};\quad
	\vartilde{\alpha}^* = 2.35023;\quad
	\vartilde{R}^* = \SI{11.122}{\kilo\meter};
\end{equation}
where the true radius has been obtained by equation
\eqref{eq:BBlaw}. Then, the flux can be studied
as it were composed by one non-oscillating
effective species $\vartilde{\nu}$. Again, we
perform three different analyses, taking the
priors on $\vartilde{\alpha}$ listed in table
\ref{tab:param} and restricting
ourselves to the $D=\SI{10}{\kilo\parsec}$ case.

\begin{figure}[t]
	\centering
	\subfloat[][$\vartilde{\alpha}$--$\vartilde{R}$
	plane projection]
	{\includegraphics[width=.35\textwidth]
	{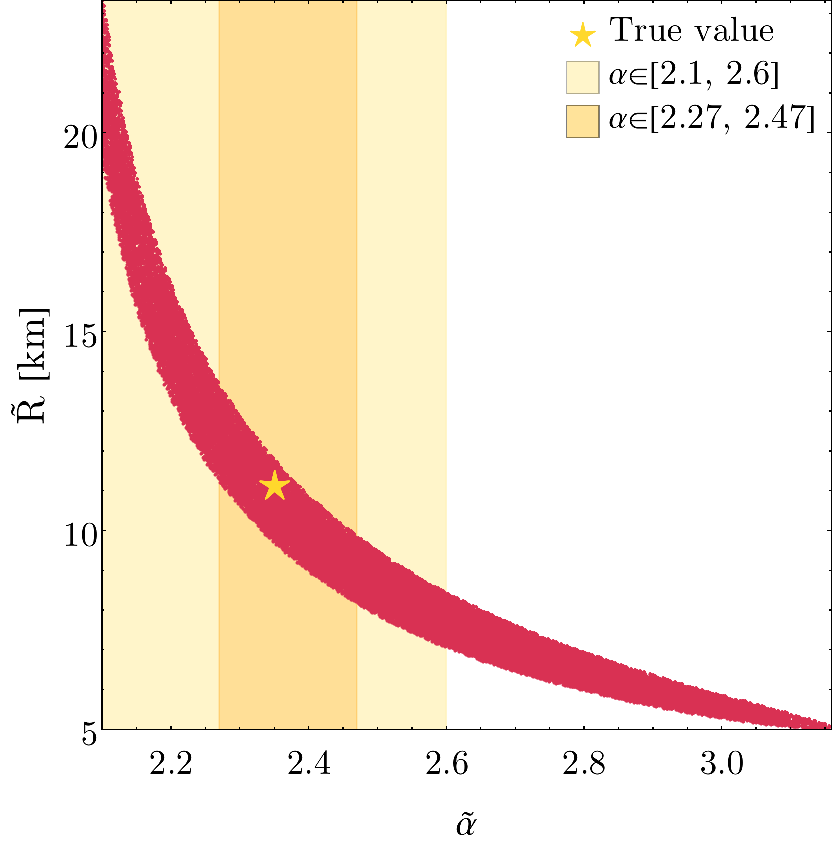}
	\label{fig:Sc1s}}
	\subfloat[][Radius reconstruction]
	{\includegraphics[width=.55\textwidth]
	{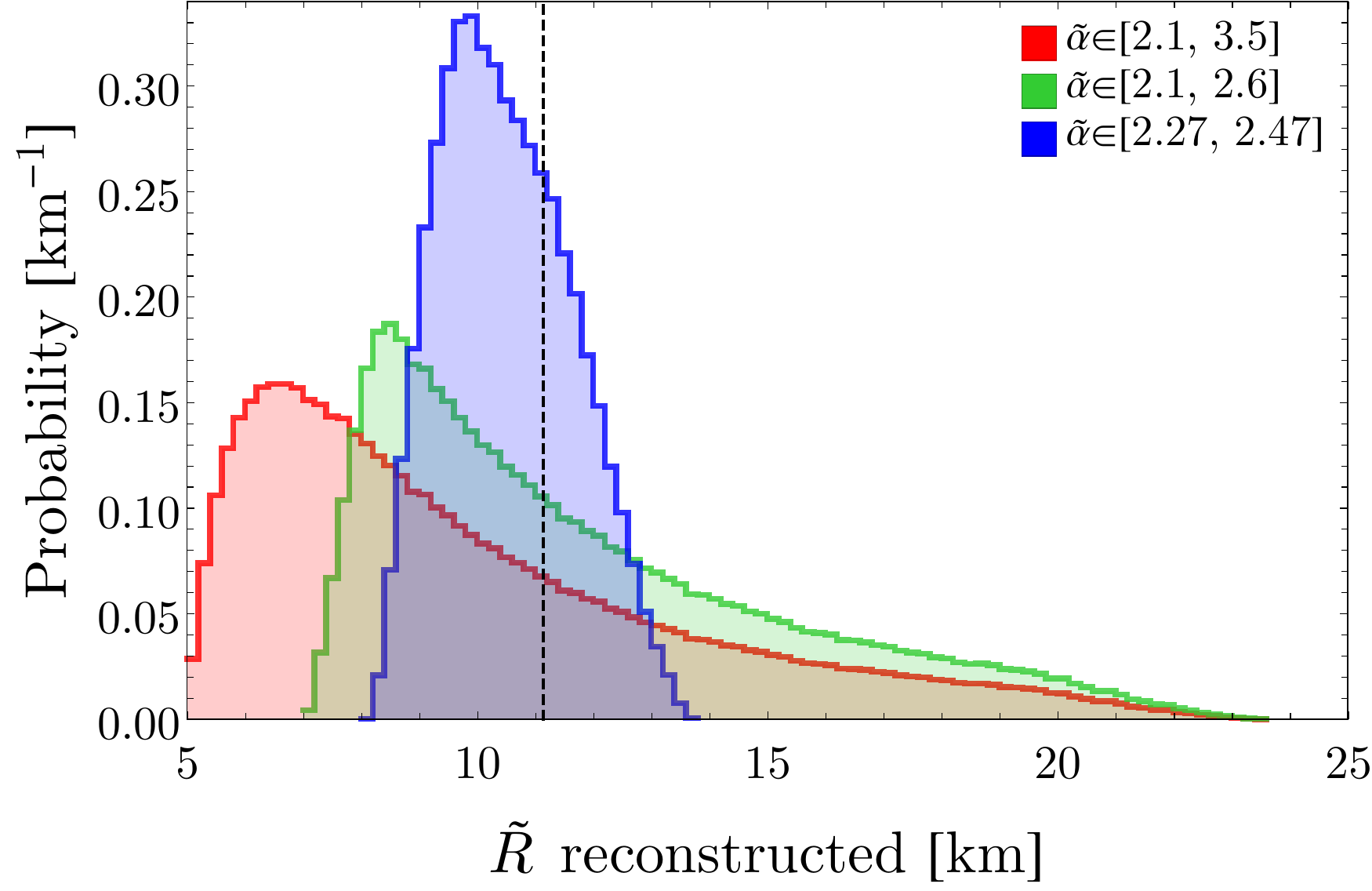}
	\label{fig:HR1s}}
	\caption{Panel \protect\subref{fig:Sc1s}
	shows the projection of 40k accepted points at
	$3\sigma$ CL onto the $\vartilde{\alpha}
	$--$\vartilde{R}$ plane. The star marks the
	true values, while the bands highlight the
	different assumptions on the prior of $\alpha$.
	Panel \protect\subref{fig:HR1s} shows the
	distributions of the reconstructed radii in the
	three different analyses, characterized by a
	different prior of $\alpha$. The dashed line
	marks the true value.
	}
	\label{fig:Plot1s}
\end{figure}

For completeness, the results are shown in table
\ref{tab:risu2}. As one can see, the normalization
$\vartilde{\mathcal{E}}$ and the mean $\langle
\vartilde{E}\rangle$ of the distribution are
measured with great accuracy, namely $\sim 4\% $
and $\sim 3\%$ in the less constrained analysis.
This precision does not improve much with the
most restrictive $\vartilde{\alpha}$ prior. On the
other hand, the spectral shape is not known at all,
with a distribution of the pinching almost flat in
the prior. This of course propagates to the radius.
Figure \ref{fig:Sc1s} shows the projection onto the
$\vartilde{\alpha}$--$\vartilde{R}$ plane of 40k
accepted points at $3\sigma$ CL. Figure
\ref{fig:HR1s} reports the distribution of the
reconstructed radius, assuming different priors on
$\vartilde{\alpha}$. As one can see, the parameter
region is thinner with respect to figure
\ref{fig:Scatter3}, but the correlation is still
present. This means that, in the most conservative
case, the emission radius is determined within
$\sim 40\%$; while it can be reduced up to $10\%$,
but with an aggressive prior. 

\begin{table}[t]
	\centering
	\begin{tabular}{|c|c|ccc|ccc|}
		\cmidrule{3-8}
		\multicolumn{1}{c}{~}
		&\multicolumn{1}{c|}{\Tstrut~\Bstrut}
		&\multicolumn{3}{c|}{$R$--\textsc{prior}}
		&\multicolumn{3}{c|}{%
		$R$--\textsc{prior}\texttt{+}}\\
		\cmidrule{2-8}
		\multicolumn{1}{c|}{\Tstrut~}
		&\multicolumn{1}{c|}{true\Bstrut}
		&\multicolumn{1}{c}{Mean}
		&\multicolumn{1}{c}{SD}
		&\multicolumn{1}{c|}{Acc}
		&\multicolumn{1}{c}{Mean}
		&\multicolumn{1}{c}{SD}
		&\multicolumn{1}{c|}{Acc}\\
		\hline
		\multirow{2}{*}{$\nu_{\mathrm{e}}$}
		& \multirow{2}{*}{2.39}
		& 2.79 & 0.39 & 14.1 & 2.78 & 0.39 & 14.0\\
		&&\skpc{2.83} & \skpc{0.37} & \skpc{13.1}
		& \skpc{2.83} & \skpc{0.37} & \skpc{13.1}\\
		\hline
		\multirow{2}{*}{$\overline{\nu}_
		{\mathrm{e}}$}
		& \multirow{2}{*}{2.28}
		& 2.33 & 0.16 & 6.73 & 2.27 & 0.1 & 0.55\\
		&& \skpc{2.33} & \skpc{0.11} & \skpc{4.81}
		& \skpc{2.29} & \skpc{0.053} & \skpc{2.3}\\
		\hline
		\multirow{2}{*}{$\nu_x$}
		& \multirow{2}{*}{2.46}
		& 2.58 & 0.27 & 10.3 & 2.5 & 0.19 & 7.45\\
		&& \skpc{2.51} & \skpc{0.17} & \skpc{6.96}
		& \skpc{2.45} & \skpc{0.09} & \skpc{3.67}\\
		\hline
	\end{tabular}
	\caption{Results of the $\alpha$
	parameter reconstruction in Hyper-Kamiokande,
	for the three neutrino species, from the
	time-window $6\div\SI{10}{\second}$, combining
	IBD and ES signal. The analysis has been
	performed with the default $\alpha$ prior,
	keeping only the points whose radii are inside
	the radius prior. The two blocks named
	``$R$--prior'' and ``$R$--prior\texttt{+}''
	refer to a $R$ prior of $[8,\,16]$
	\si{\kilo\meter} and $[10.2,\,13.1]$
	\si{\kilo\meter} respectively. The top rows
	present the $D=\SI{10}{\kilo\parsec}$
	analysis, while the quantities in parentheses
	are the result of a $D=\SI{2}{\kilo\parsec}$
	explosion. For each $\alpha_i$ we give its
	true value, the mean
	of the parameter distribution, its standard 
	deviation (SD) and the overall accuracy.}
	\label{tab:risu3}
\end{table}

\begin{figure}[t]
\begin{centering}
	\includegraphics[width=.9\textwidth]
	{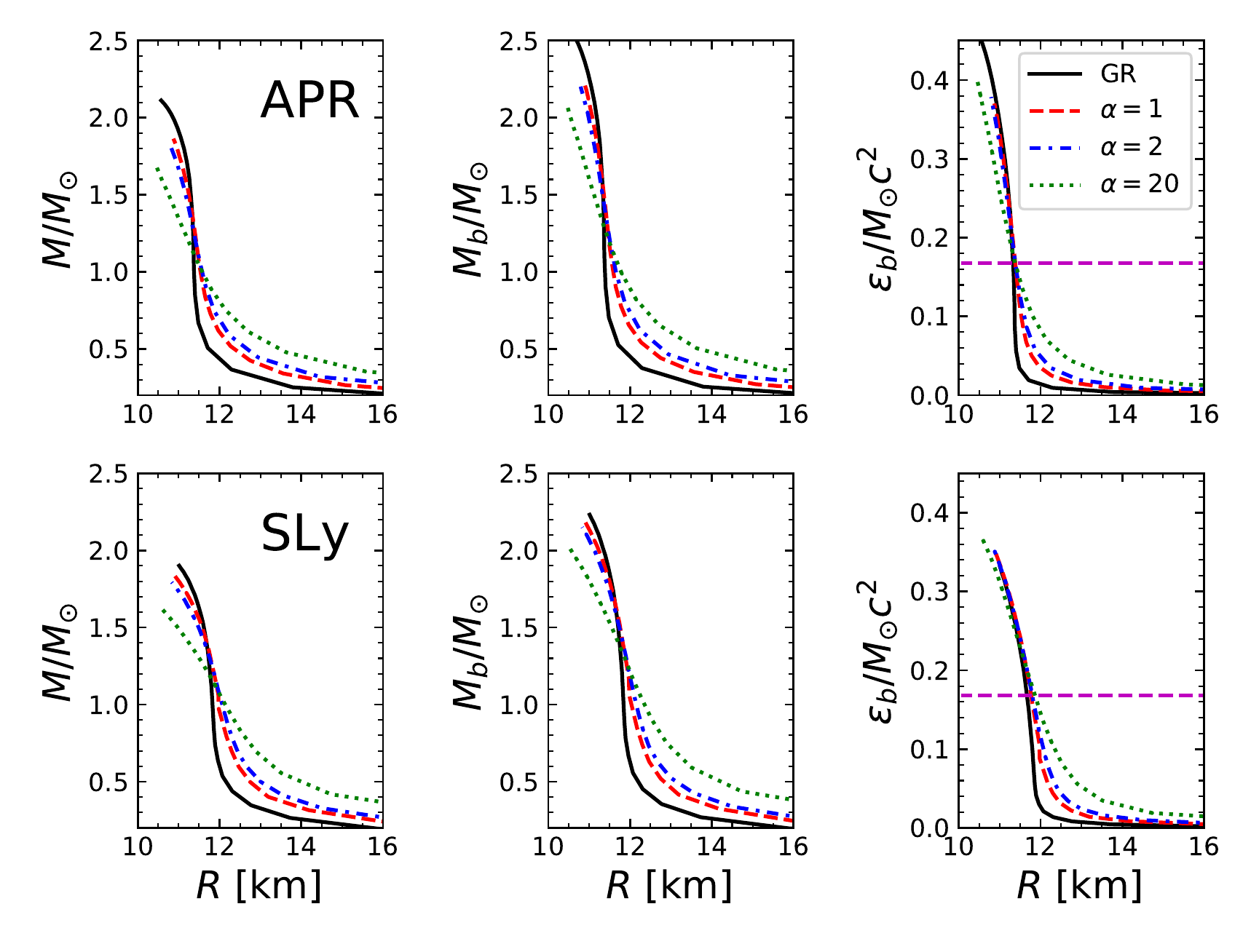}
	\caption{Predictions for macroscopic properties
	of a neutron star, for the two equations of
	state APR (upper) and SLy (lower figures). The
	quantities on the $y$-axis --- neutron star
	mass (left), baryonic mass (middle) and
	gravitational binding energy (right figures)
	--- are plotted as a function of the neutron
	star radius. $M_{\odot}$ is the Sun mass. The
	results correspond either to predictions in
	general relativity (GR) or for $f(R)$ theories
	with $f(R) = R + \alpha R^2$ and three
	different $\alpha$ values (see text). To guide
	the eye, we plot the value $\SI{3e53}{\erg}$ 
	(the magenta dashed lines) in the third
	panels. }
	\label{fig:HistoSj}
	\end{centering}
\end{figure}

Neutron star masses and radii depend on the neutron
star equation of state and also could be modified
in extended theories of gravity. Future
measurements with X-rays and gravitational waves
will obtain tight constraints on the mass and
radius relation of cold neutron stars and on the
radius itself. Gravitational waves observations
might discover extended theories of gravity. With
all the {\it caveats} of the present analysis, we
would like to show the sensitivity of the
gravitational binding energy of the nascent neutron
star to theories beyond general relativity such as
the so-called $f(R)$ theories
\cite{Capozziello:2015yza}. Figure
\ref{fig:HistoSj} shows the gravitational
mass, the baryonic mass as well as the
gravitational binding energy as a function of the
neutron star radius. Results are shown both for 
general relativity as well as for
$f(R) = R + \alpha R^2$ extended theory of gravity
for three different values of $\alpha$ (see
appendix A.2 for the corresponding equations).
While masses and radii differ significantly from
the general relativity predictions for large
$\alpha$ values as expected, the gravitational
binding energy turns out to have much smaller
sensitivity to such extended theories of gravity
even for $\alpha$ as large as 20.
The results are plotted for two different equations
of state, i.e.\ APR and Sly \cite{Oertel:2016bki}.

\subsection{Inverting the perspective:
constraining the effective pinching}

Our results show that the reconstructed radii
vary significantly with $\alpha$. Conversely,
assuming a prior knowledge on the radius could help
in the determination of the pinching  parameter
$\alpha$. Indeed, we can take the extracted
points in the IBD$+$ES three flavor analysis (with
broad $\alpha$) and analyze only those whose radii
are inside a defined physical prior.
Current neutron star equations of state, compatible
with neutron star observations, predict $R \in 
[8, 16]$ \si{\kilo\meter}. This is our first prior
for the neutrinosphere, called ``$R$--prior''.
On the other hand,
we can also assume an aggressive prior about the
radius knowledge, i.e.\ $R \in 
[10.2, 13.1]$ \si{\kilo\meter} which could result
from independent measurements. It will be called
``$R$--prior\texttt{+}''. The results for the
pinching reconstruction for the two priors
are given in table \ref{tab:risu3}.

Thanks to the correlation between $\alpha$ and $R$,
under the $R$--prior case, an accuracy of $7\%$ and
$10\%$ can be achieved on the pinching parameters 
of $\overline{\nu}_{\mathrm{e}}$ and $\nu_x$
respectively (for a supernova at
\SI{10}{\kilo\parsec}).
With the more aggressive $R$--prior\texttt{+},
the accuracy improves to $\sim 1\%$ and $\sim 7\%$
for the $\overline{\nu}_{\mathrm{e}}$  and $\nu_x$
species respectively. As expected, the
determination of $\alpha$ for the
$\nu_{\mathrm{e}}$ species does not improve.

Concerning the other flux parameters, the accuracy
of the reconstructed values is very similar for
both priors: the accuracy on the emitted energies
for $\overline{\nu}_{\mathrm{e}}$ 
and $\nu_x$ species is $\sim 20\%$;
the accuracy on their mean energies is
$\sim 7\div 8\%$.
These values are similar to the ones reported in
ref.\ \cite{GalloRosso:2017mdz} for
Super-Kamiokande, although we remark that
the emitted energy $\mathcal{E}_{\APnue}$ was there
determined with an accuracy of $\sim 10\%$.

\section{Conclusions}
\label{sec:quattro}
We have explored the possibility to reconstruct the
radius of the newly formed neutron star in a
core-collapse supernova explosion through its
neutrino time signal. To this aim, we have
performed a nine-degrees of freedom likelihood
analyses, considering the total energy, the average
energy and the pinching parameter for the
three neutrino species. These characterize th
$\nu_e$, $\overline{\nu}_{\mathrm{e}}$ and $\nu_x$
fluences which are assumed to be black-body spectra
equipped with pinching to account for the
deviations from  thermal distributions. The
neutrino time signal is adopted from simulations by
Roberts and Reddy. The interval of time chosen for
the analysis, between 6 and 10 seconds, is
motivated by the fact that the neutron star and 
neutrinosphere radii are approximatively time
independent over this range which justifies the use
of  fluences, instead of the time-dependent
neutrino signal \cite{Roberts:2016rsf}.  

We have combined inverse beta decay and elastic
scattering in a water Cherenkov detector of the
size of the future Hyper-Kamiokande, assuming 
ideal detector performance --- namely, an
efficiency approaching $100 \%$.
The numerical results have unravelled a tight
correlation between the pinching parameters and the
reconstructed neutron star radius. 
We performed three analysis, ranging from a very
conservative to a rather aggressive one with
respect to the prior choices of the pinching
parameter. In the most conservative one, the
neutrinosphere radius could
be determined only with a precision of  $35\%$
($56 \%$) for a supernova at \SI{2}{\kilo\parsec}
(\SI{10}{\kilo\parsec}) for $\overline{\nu}
_{\mathrm{e}}$; whereas with the most
aggressive prior the precision could improve
to  $12 \%$  ($26 \%$) for a supernova at
\SI{2}{\kilo\parsec} (\SI{10}{\kilo\parsec}).
A similar accuracy is obtained for $\nu_x$.
While these results might be sufficient to probe
(exclude) the cases of oscillations into mirror
\cite{m2} or pseudo-Dirac neutrinos \cite{pd2},
they indicate the difficulty in determining the
neutron star radius precisely through neutrino
measurements. This conclusion is not altered if
statistics is increased since both the
\SI{10}{\kilo\parsec} and the \SI{2}{\kilo\parsec}
supernova results show similar quantitative trends.
Our work indicates the importance of getting
precise knowledge on the width of the quasi-thermal
neutrino distribution from core-collapse supernova
simulations.

One should consider that the neutrino time signal
is directly related to the neutrinosphere radii
and is a unique observable to get 
information about them. 
In order to successfully perform the type of
analysis described here, and to interpret the
results in connection with the neutron star radius,
a systematic study will
be required to quantitatively connect the
neutrinosphere radii to the radius of the neutron
star formed during the explosion which are thought
to be within $10\div 15\%$ of each other for the
time interval of interest. 

By inverting the perspective, we have performed two
more analyses: a) implementing in the likelihoods
that the neutron star radius (and consequently the
radius of the neutrinosphere) varies in the range
$8\div\SI{10}{\kilo\meter}$; 
b) assuming that future measurements will narrow
its current uncertainty down to $10\%$. In case a)
the pinching parameters of the neutrino fluences
can be determined with an accuracy of $5\%$
($7\%$) at \SI{2}{\kilo\parsec} (\SI{10}
{\kilo\parsec}) for $\overline{\nu}_{\mathrm{e}}$
and $7 \%$ ($10 \%$) at \SI{2}{\kilo\parsec}
(\SI{10}{\kilo\parsec}) for $\nu_x$. In case b)
the pinching parameters of the neutrino
fluences can be determined with an accuracy of 
$1 \%$ ($2 \%$) at \SI{2}{\kilo\parsec}
(\SI{10}{\kilo\parsec}) for  $\overline{\nu}
_{\mathrm{e}}$ and $4 \%$ ($7 \%$) at
\SI{2}{\kilo\parsec} (\SI{10}{\kilo\parsec}) for
$\nu_x$. The other parameters of the fluences can
be determined with good precision as well. 
Likewise, for $\nu_{\mathrm{e}}$ the reconstruction
of the corresponding pinching is approximately with
a precision of $14 \%$ while the determination of
the fluence would require the inclusion of
$\nu_{\mathrm{e}}$ sensitive detection
channels.

\section*{Acknowledgments}
\addcontentsline{toc}{section}{Acknowledgments}

The authors are grateful to S.\ Capozziello for
useful discussions. M.C.\ Volpe acknowledges
financial support  from ``Gravitation et physique
fondamentale'' (GPHYS) of the {\it Observatoire de
Paris}.

\appendix

\section{Tolman-Oppenheimer-Volkoff equations}
In order to obtain the mass-radius relation for
neutron stars we have solved the
Tolman-Oppenheimer-Volkoff (TOV) equations which
govern the physics of the matter-geometry in
spherically symmetric space-time
\cite{Oppenheimer:1939ne}.
We have considered both general
theory of relativity and extended theories of
gravity, in particular the so-called $f(R)$
theories. We also comment on how to deal with the
problem numerically so that the interested reader
could reproduce the plots in figure
\ref{fig:HistoR3}.

\subsection{General relativity case}
The geometry of a spherically symmetric space-time
could be described by the metric specified by
\begin{equation}
\df{s}^2 = e^{2w} c^2 \df t^2 - e^{2\lambda}
\df r^2 - r^2(\df\theta^2 + \sin^2\theta \df\phi^2),
\end{equation}
where $r$ denotes the radial variable and
$w(r)$ and $\lambda(r)$ are functions of $r$. 
From the Einstein's equation one can derive the TOV
equations in which the three degrees of freedom
$p$, $\lambda$ and $w$ are then governed by 
\begin{align}
\frac{\df{p}}{\df{r}} &= -(\rho c^2 + p)
\frac{\df{w}}{\df{r}}, \\
\frac{\df\lambda}{\df r} &=\frac{4\pi r G\rho}{c^2}
e^{2\lambda} - \frac{e^2\lambda}{2r}+\frac{1}{2r},
\end{align}
and
\begin{equation}
\frac{\df w}{\df r} = \frac{4\pi r G p}{c^4}
e^{2\lambda} + \frac{e^2\lambda}{2r} -
\frac{1}{2r},
\end{equation}
where $G$ is the constant of gravity and the
energy density $\rho$, in principle, is related to
the pressure $p$ through the equation of state
(EOS).

To solve the TOV equations in GR numerically, one
should note that the TOV equations for $p$ and
$\lambda$ could be written in a form which is
explicitly decoupled from $w$. This means that to
integrate them, it is only required to provide two
initial values. One for $\lambda$ and one for $p$,
namely $p(r=0) = p(\rho_c)$ and $\lambda(r=0)
= 0$ with $\rho_c$ being the value of the energy
density at the center of the neutron star. 

\subsection[TOV equations in  $f(R)$ gravity]{
\boldmath TOV equations in  $f(R)$ gravity}
In modified theories of gravity, the equation
relating the matter to the geometry could be
different from the Einstein equation. In
particular, in $f(R)$ theory of gravity (in the
metric formalism), it could be written as
\cite{Capozziello:2015yza} 
\begin{equation}
\frac{\df f(R)}{\df{R}}R_{\mu\nu} - \frac{1}{2}
	f(R)g_{\mu\nu} - \left[\nabla_\mu \nabla_\nu -
 	g_{\mu\nu} \Box\right]\frac{\df f(R)}{\df R}
 	= \frac{8\pi G}{c^4}T_{\mu\nu}, 
\end{equation}
where $R$ is the Ricci scalar and $T_{\mu\nu} $
is the energy momentum tensor.
In spherically symmetric space time, one can find
the generalized TOV
equations in $f(R)$ gravity as
\begin{align}
	\frac{\df p}{\df r} &= -(\rho c^2 + p)
	\frac{\df w}{\df r}, \\
	\frac{\df\lambda}{\df r} &=
	\frac{8\pi r G
	\rho e^{2\lambda}}{c^2(2f'_R + rR' f''_R)}+
	\frac{e^{2\lambda} [(r^2R-2)f'_R - fr^2]}
	{2r(2f'_R + rR'f''_R)}
	+\frac{f'_R +
	r[f''_R(2R'+rR'') + rR'^2 f'''_R]}{r(2f'_R +
	rR'f''_R)}
\end{align}
and
\begin{equation}
\frac{\df w}{\df r} =  \frac{8\pi r G p
e^{2\lambda}}{c^4(2f'_R + rR' f''_R)} +
\frac{e^{2\lambda} [f r^2 + (2-r^2R)f'_R ] -
2(f'_R + 2rR'f''_R)}{2r(2f'_R + rR'f''_R)}, 
\end{equation}
where the prime $'$ denotes the derivative with
respect to $r$ while $f'_R$, $f''_R$ and $f'''_R$
are the first, second and third order derivatives
of $f(R)$ with respect to $R$. Unlike the case of
GR where $R$ is determined statically and
explicitly in terms of $\rho$ and $p$
\begin{equation}
R =  \frac{8\pi G}{c^4} (\rho c^2 - 3p), 
\end{equation}
in modified gravity, the Ricci scalar could be
completely dynamic which adds to the complexity of
solving the TOV equations in these theories. 
In particular, in $f(R)$ gravity one has
\begin{equation}
\frac{\diff{2} R}{\df r^2}  = \frac{1}{3f''_R}
\left\{ -\frac{3f''_R R'}{r}(rw'-r\lambda ' +2)
-3f'''_R R'^2 + e^{2\lambda}(R f'_R -2f) +
\frac{8\pi G e^{2\lambda}}{c^4} (\rho c^2 - 3p)
\right\}.
\end{equation}
The generalized TOV equations can be derived
for $f(R) = R + \alpha R^2$ gravity as
\begin{align}
\frac{\df p}{\df r} &= -(\rho c^2 + p) 
\frac{\df w}{\df r}, \\
\frac{\df\lambda}{\df r} &=  \frac{4\pi r G \rho
e^{2\lambda}}{c^2(1+2\alpha R + \alpha r R')} 
+ \frac{  e^{2\lambda}(\alpha r^2 R^2 - 4 \alpha
R - 2) + 2  + 4\alpha r^2 R'' + 8\alpha r R' +
4\alpha R}{4r  + 8\alpha r + 4 \alpha r^2 R'},\\
\frac{\df w}{\df r} &=  \frac{4\pi r G p
e^{2\lambda}}{c^4(1+2\alpha R+ \alpha r R')} + 
\frac{ e^{2\lambda}(-\alpha r^2 R^2 + 4
\alpha R + 2) - 2 - 4\alpha R - 8\alpha r R'}{4r 
+ 8\alpha r R + 4 \alpha r^2 R'},
\end{align}
and 
\begin{equation}
\frac{\diff{2} R}{\df r^2} = \frac{1}{6 c^4 r
\alpha} \left\{ e^{2\lambda}( 8\pi r G (\rho c^2
- 3p) -c^4 r R ) - 6c^4\alpha R'(2+rw'-r\lambda ')
\right\}.
\end{equation}
It should be noted that to avoid the ghost
instabilities, one could only opt for negative
values of $\alpha$, i.e.\ $\alpha < 0$
\cite{DeFelice:2006pg}.

Given TOV equations in $f(R)$ gravity, one can
integrate them numerically. Here, there are four
coupled differential equations for $p$, $\lambda$,
$w$ and $R$ which are required to be solved
simultaneously. Moreover, one should  note that the
differential equation for $R$ is second order.
To solve these equations numerically, one 
should know the initial (central) values of $p$,
$\lambda$, $w$, $R$ and $R'$. The natural choices
could be  $p(r=0) = p(\rho_c)$, $\lambda(r=0) = 0$
and $R'(r=0) = 0$. Moreover, the generalized 
TOV equations are linear in $w$ as in GR and one 
does not need to know the central value of 
$w$ beforehand. What should be done is just to
take an initial value for $w_c$ and then correct
it after solving the equations so that it matches
the correct boundary value at infinity. 
However, there is still a big difference between
$f(R)$ theories and GR in dealing with TOV
equations numerically. In the case of $f(R)$
gravity, one does not know beforehand the value of $R_c =
R(r=0)$ 
which is needed to integrate the TOV
equations. To overcome this difficulty, the
strategy we chose was that we solved the TOV
equations by using the bisection method for the 
range $R_c = R_{GR}/10$ to $R_c = 10
R_{GR}$.\footnote{We had to take much larger
range of $R_c$ for very large values of $\alpha$.}
It turns out that (at least for $R^2$
gravity)\footnote{We noted that this could fail
for some other $f(R)$ gravity.} the value of $R_c$
which could result in a physical behavior for
$\lambda$ and $R$ is very specific and only for a
very narrow range of $R_c$ we observed the 
physically expected asymptotic behavior for this
quantities. For other values of $R_c$, for
example,  $\lambda$ could grow/drop monotonously
when $r \rightarrow \infty$ instead of the 
expected asymptotic behavior $\lambda
\rightarrow 0$.

\end{document}